\documentclass[12pt]{article}

\usepackage[round]{natbib}
\usepackage[utf8]{inputenc}
\usepackage[all]{xy}
\usepackage{amssymb}
\usepackage{amsmath}
\usepackage{amsthm}
\usepackage{csquotes}
\usepackage{subfig}
\usepackage{enumerate}
\usepackage{graphicx}
\usepackage{enumerate}
\usepackage{color}
\usepackage{mwe}
\usepackage{appendix}
\usepackage{centernot}
\usepackage[linesnumbered,ruled,vlined]{algorithm2e}
\usepackage{pgfplots}
\usepackage{rotating}
\usepackage{verbatim}
\usepackage{authblk}
\usepackage{mathtools}
\usepackage{pdfpages}
\usepackage{import}

\usepackage{soul}
\usepackage{xcolor}
\def\@centernot#1#2{%
  \mathrel{%
    \rlap{%
      \settowidth\dimen@{$\m@th#1{#2}$}%
      \kern.5\dimen@
      \settowidth\dimen@{$\m@th#1=$}%
      \kern-.5\dimen@
      $\m@th#1\not$%
    }%
    {#2}%
  }%
}
\makeatother

\newcommand{\blind}{1}

\newcommand{\independent}{\perp\mkern-9.5mu\perp}

\newcommand{\sign}{\text{sign}}

\usepackage{tikz}
\usetikzlibrary{arrows.meta,shapes}
\usetikzlibrary{arrows,shapes.arrows,shapes.geometric,shapes.multipart,
decorations.pathmorphing,positioning,shapes.swigs,}

\newtheorem{lemma}{Lemma}
\newtheorem{proposition}{Proposition}
\newtheorem{theorem}{Theorem}
\newtheorem{corollary}{Corollary}
\theoremstyle{definition}
\newtheorem{assumption}{Assumption}
\newtheorem{definition}{Definition}
\newtheorem{example}{Example} 
\theoremstyle{remark}
\newtheorem{remark}{Remark}

\usepackage{xr}
\makeatletter
\newcommand*{\addFileDependency}[1]{
\typeout{(#1)}
\@addtofilelist{#1}
\IfFileExists{#1}{}{\typeout{No file #1.}}
}\makeatother
\newcommand*{\myexternaldocument}[1]{%
\externaldocument{#1}%
\addFileDependency{#1.tex}%
\addFileDependency{#1.aux}%
}
\myexternaldocument{appendix}

\MakeOuterQuote{"}\EnableQuotes
\DeclareUnicodeCharacter{00A0}{ }

\pgfplotsset{compat=1.18} 

\begin{document}

\def\spacingset#1{\renewcommand{\baselinestretch}%
{#1}\small\normalsize} \spacingset{1}


\if1\blind
{
  \title{\bf Improved bounds and inference on optimal regimes}
  \author[1]{Julien D. Laurendeau \thanks{
    The authors gratefully acknowledge support from the Swiss National Science Foundation }\hspace{.2cm}\thanks{julien.laurendeau@epfl.ch}}
  \author[1]{Aaron L. Sarvet \thanks{aaron.sarvet@epfl.ch}}
  \author[1]{Mats J. Stensrud \thanks{mats.stensrud@epfl.ch}}
    \affil[1]{Institute of Mathematics, Ecole Polytechnique Fédérale de Lausanne, Station 8, 1015 Lausanne, Switzerland}
    \date{}
  \maketitle
} \fi

\if0\blind
{
  \bigskip
  \bigskip
  \bigskip
  \begin{center}
    {\LARGE\bf Improved bounds and inference on optimal regimes}
\end{center}
  \medskip
} \fi

\begin{abstract}
Point identification of causal effects requires strong assumptions that are unreasonable in many practical settings. However, informative bounds on these effects can often be derived under plausible assumptions. Even when these bounds are wide or cover null effects, they can guide practical decisions based on formal decision theoretic criteria. Here we derive new results on optimal treatment regimes in settings where the effect of interest is bounded. These results are driven by consideration of superoptimal regimes; we define regimes that leverage an individual's natural treatment value, which is typically ignored in the existing literature. We obtain (sharp) bounds for the value function of superoptimal regimes, and provide performance guarantees relative to conventional optimal regimes. As a case study, we consider a commonly studied Marginal Sensitivity Model and illustrate that the superoptimal regime can be identified when conventional optimal regimes are not. We similarly illustrate this property in an instrumental variable setting. Finally, we derive efficient estimators for upper and lower bounds on the superoptimal value in instrumental variable settings, building on recent results on covariate adjusted Balke-Pearl bounds. These estimators are applied to study the effect of prompt ICU admission on survival.
\end{abstract}




\noindent%
{\it Keywords: Causal inference, Partial identification, Optimal treatment regimes, Semiparametric estimation} 

\spacingset{1.9} 
\section{Introduction} 
A precision medicine system assigns treatments to patients that are optimally adapted to their personal characteristics.  Similarly, a wide range of industries, e.g.\ in the technology sector, strive to target their products to individuals. Because individual decisions, for example about medical treatments, are selected based on their anticipated effects, identification of individualized regimes is a causal inference task. Motivated by the broad interest in this domain, the causal literature on optimal regimes has flourished in the last decades \citep{murphy2003optimal, robins2004optimal,tsiatis2019dynamic,kosorok2021introduction}. However, most of the existing results rely on assumptions about no unmeasured confounding \citep{tsiatis2019dynamic}[for an overview]. These assumptions are certainly convenient, as they might allow point identification of optimal regimes and the accompanying value function. Yet, such assumptions are often implausible in practice. In particular, many health care providers, such as medical doctors, have access to patient information that is not recorded in observational datasets. For example, doctors are receptive to visual and auditory cues that are hard to record in computers \citep{hamerman1999toward}. If such information affects doctors' treatment decisions and is also associated with patient outcomes, then there is likely unmeasured confounding in the data.  

The issue of unmeasured confounding is routinely cited as a severe problem for valid causal inference and has fostered much debate and research. Recent work has considered point identification and estimation of optimal regimes in the presence of unmeasured treatment-outcome confounding, in particular using instrumental variable (IV) methods \citep{qiu2021optimal,han2021comment,cui2021necessary,cui2021semiparametric}. While these methods permit identification in the presence of unmeasured confounding, alternative assumptions are required. Yet the alternative assumptions may be equally challenging to justify \citep{hernan2006instruments}, or may even contradict investigators' baseline hypotheses underlying their research, for example, that there is meaningful heterogeneity in patients' responses to treatment. 

If the investigator is unwilling to make assumptions about unmeasured confounding, exclusion restrictions or effect homogeneity, point identification of optimal regimes, and counterfactual outcome values under these regimes, is often difficult \citep{cui2021necessary}. However, partial identification results, which provide bounds, can still be obtained under assumptions that the analysts deem to be plausible \citep{robins_analysis_1989,Manski1990,BalkePearl1997,manski_monotone_2000,richardson_ace_2014,swanson2018partial,cui2021necessary,pu2021estimating,han_optimal_2023,han_computational_2023}. These bounds might be wide and cover the null effect, reflecting the uncertainty that remains after fully leveraging knowledge of the observed distribution and all external assumptions. Nevertheless, the use of decision criteria can be justified by formal decision theoretic results \citep{manski_identification_2000, manski_treatment_2002,stoye_essays_2005, Cui2021Individualized}.  

\subsection{Our contribution}
We derive results on the identification and estimation of bounds on optimal regimes in the presence of unmeasured confounding. Unlike existing work, the bounds we consider leverage the so-called natural treatment value \citep{richardson2013single}, representing the treatment an individual would choose or be given absent of its assignment in an experiment. The use of this additional variable gives superoptimality properties relative to classical optimal regimes \citep{StensrudSarvet2022}. We further describe settings where the treatment regime of interest is point identified when using the natural treatment value, even if conventional optimal regime methods only give bounds. Such settings comprise a non-pathological subspace of data generating mechanisms, as illustrated by our examples. More broadly, we describe algebraic relationships, clarifying how bounds on superoptimal regimes can be obtained from existing results and how the width of these bounds are related. We give a simple algorithm for obtaining bounds on superoptimal regimes and their value functions, based on conventional bounds on conditional average treatment effects. As a further illustration, we examine performance guarantees of previous proposals \citep{kallus2018interval,Yadlowsky2022}, and show that the superoptimal regimes can improve these guarantees. We then give detailed results on identification and estimation of bounds under the binary IV model \citep{BalkePearl1997,swanson2018partial}. Finally, we use the methods to study the effect of prompt ICU admission on survival \citep{harris2015delay, keele2020stronger, StensrudSarvet2022}, using the number of available ICU beds as an instrument. Two additional data analyzes are presented in the online Appendices.


\subsection{Relation to previous work}
\label{sec: rel previous work}
This article builds on results on dynamic treatment regimes that are functions of the natural treatment value, see, e.g.,  \citet{robins2007causal}, \citet{geneletti2011att}, \citet{munoz_population_2012} and \citet{young2014identification}. Similarly, interventions that are functions of the natural treatment value appear in works on so-called "modified treatment policies" (MTPs), that are also defined by functions of the natural treatment value \citep{Haneuse_estimation_2013, Diaz_causal_mediation_2020, Diaz2021lmtp}. These works predominantly considered multi-level treatments in longitudinal settings with no unmeasured confounding. Furthermore, these works did not give results on optimal regimes. In contrast, we consider optimal regimes that are deterministic functions of a binary treatment and potentially high-dimensional covariates. \citet{StensrudSarvet2022} studied treatment regimes that use binary natural treatment values to get new performance guarantees for optimal regimes.  Unlike \cite{StensrudSarvet2022}, we present results in the setting where the value functions and regimes are not point identified. This requires different definitions of optimality, partial identification results, conditions that guarantee sharpness, explicit criteria for decision making, and new estimators. 

Our results are also different from previous work on partial identification of optimal regimes based on conditional average treatment effects (CATEs) \citep{Cui2021Individualized,kallus_minimax-optimal_2021,pu2021estimating}, because we consider decision criteria that are functions of the natural treatment value. Using the natural treatment value, superoptimal regimes are always better than, or as good as, the previously considered optimal regimes. Heuristically, superoptimal regimes use more information than what is encoded in the conventional measured pretreatment covariates, which allows them to outperform optimal regimes without necessarily imposing additional assumptions. 

\citet{pu2021estimating} introduced a weighted misclassification risk function that adds a penalty corresponding to the worst case outcome for each wrong decision. Specifically, if we incorrectly decide to treat an individual, the penalty is the absolute value of the lower bound of the conditional average treatment effect. If we incorrectly decide not to treat, the penalty is the absolute value of the upper bound of the conditional average treatment effect. \cite{pu2021estimating} seeked to minimize the average penalty and obtain an "IV-optimal regime". \cite{Cui2021Individualized} showed that the IV-optimal regime is included in a broader class of regimes that minimize a certain weighted risk function. We consider some of the decision criteria from \cite{Cui2021Individualized}, adapted to our setting, including the IV-optimal regime, corresponding to the ``opportunistic'' criterion \citep{Cui2021Individualized}. 

 \cite{kallus_minimax-optimal_2021} considered the setting where the sign of the CATEs can be deduced, e.g., from bounds. They further gave certain optimality guarantees for an optimal regime under a Marginal Sensitivity Model. The IV-optimal regimes \citep{pu2021estimating} and the decision criteria from \cite{Cui2021Individualized} will always match the optimal regime from \cite{kallus_minimax-optimal_2021} when the sign of these CATEs can be deduced from bounds. Thus, the decision criteria that we will consider are not improved using the procedure in \cite{kallus_minimax-optimal_2021}. We compare a decision criterion based on their procedure to other decision criteria and to point identification methods from \citet{StensrudSarvet2022} in our analysis in Section \ref{sec: ICU example} with the healthcare decision criterion from \cite{Cui2021Individualized}, which is equivalent to using the procedure in \cite{kallus_minimax-optimal_2021} when "no treatment" is defined as baseline regime.


\section{Data structure, estimands and basic properties}
\label{sec: dat_struct}
Consider a treatment variable $A\in \{0,1\}$, a pre-treatment vector $L\in\mathcal{L}$, and an outcome $Y\in\mathbb{R}$. Assume that we have $n$ i.i.d. observations of each of these variables corresponding to specific individuals in a non-experimental (observational) setting. 

We study treatment regimes $g:\{0,1\} \times \mathcal{L} \to \{0,1\}$ that assign treatment $g(a', l)$ given the observed (natural value of) treatment $A=a'$  \citep{richardson2013single} and other pre-treatment covariates $L=l$. In a slight abuse of notation, we omit the argument $a'$ from regimes that are only trivially a function of the natural treatment value, for example, as with the conventional optimal regime. We use superscripts to denote potential outcomes. For example, $Y^a$ denotes the outcome under assignment to treatment value $a$. We will also use the shorthand $Y^g$ to denote the potential outcome under a general regime $g$. Our work concerns two different types of optimal regimes:



\begin{definition}[$L$-optimal regime]
    The $L$-optimal regime $g_{\textbf{opt}}$ assigns treatment $a$ given $L = l$ as
    \begin{equation*}
     g_{\textbf{opt}}(l) \equiv \underset{a \in \{0,1\} }{\arg \max } \  \mathbb{E}(Y^{a} \vert L = l).
\end{equation*}
\end{definition}
\begin{definition}[$L$-superoptimal regime]
The $L$-superoptimal regime $g_{\textbf{sup}}$ assigns treatment $a$, given $A = a'$, $L = l$ as
\begin{equation*}
    g_{\textbf{sup}}(a',l) \equiv \underset{a \in \{0,1\} }{\arg \max } \  \mathbb{E}(Y^{a} \mid A=a', L = l).
\end{equation*}
\label{def: LA opt regime}
\end{definition}
When convenient, we refer to the $L$-superoptimal regime as the $(L, A)$-optimal regime. We also refer to the superoptimal treatment decision $g_{\textbf{sup}}(a',l)$, i.e., the assignment for individuals with $(L,A) = (l,a')$, as the $(l,a')$-optimal regime. For clarity, we consistently use $a'$ to refer to values of the natural treatment and $a$ to refer to a values of the assigned treatment, e.g.,  $a=g(a',l)$. When $\mathbb{E}[Y^0| L = l] = \mathbb{E}[Y^1 | L = l]$, we arbitrarily set the optimal regime to $a = 0$ as both choices are equivalent. We do the same  for the $(l,a')$-optimal regime when $\mathbb{E}[Y^0 | A = a', L = l] = \mathbb{E}[Y^1 | A = a', L = l]$.

Unlike the $L$-optimal regime, the $(L,A)$-optimal regime further uses the natural treatment value $A$, which heuristically can be interpreted as a decision maker's answer to the question "Which treatment would you choose based on the measured covariates?"
 It follows from basic decision theory that the expected outcome under the $(L,A)$-optimal regime is guaranteed to be greater than or equal to that under the $L$-optimal regime \citep{bareinboim2015bandits, StensrudSarvet2022}, because any $L$-optimal regime is also in the class of regimes that are functions of $L$ and $A$. There are several results on $L$-optimal regimes in partial identification settings \citep{manski_identification_2000,manski_treatment_2002,pu2021estimating,Cui2021Individualized,cui2021machine,kallus_minimax-optimal_2021,chen_estimating_2023, han_optimal_2023}. However, to our knowledge, $(L,A)$-optimal regimes have not been studied in these settings. Yet we will argue that $(L,A)$-optimal regimes are of considerable subject-matter interest; it is precisely when there is unmeasured confounding, and thus point identification is not possible by covariate adjustment, that $(L,A)$-optimal regimes can substantially outperform $L$-optimal regimes \citep{StensrudSarvet2022}. 

Furthermore, we will show that the bounds on the $(L,A)$-optimal regimes have desirable algebraic properties such that certain conditional regimes can be point identified, as formalized in Section \ref{sec: Decision-making criteria}, even when $L$-optimal regimes are not. These results on bounds are related to, but different from the identification results in \citet{StensrudSarvet2022}, who showed that point identification of outcomes under $(L,A)$-optimal regimes requires assumptions that, in many settings, are no stronger than those under $L$-optimal regimes. In particular, these assumptions might be considered minimal, in that they may be guaranteed by design. 

\section{Identification}
\label{sec:Identification}

\subsection{Preliminary definitions}

We will distinguish between identification of regimes and identification of expected outcomes under regimes. To make these distinctions clear, two different conditional average treatment effects (CATEs) are important: 
\begin{definition}[$l$-CATE]
    The $l$-CATE for $l \in \mathcal{L}$ is $\mathbb{E}(Y^{a = 1} - Y^{a = 0} | L = l)$.
\end{definition}
\begin{definition}[$(l,a')$-CATE]
   The $(l,a')$-CATE for $(l,a') \in \mathcal{L} \times \{0,1\}$ is \\ $\mathbb{E}(Y^{a = 1} - Y^{a = 0} | A = a', L = l)$.
\end{definition} 
We will sometimes denote the $l$-CATE and $(l,a')$-CATE as $h_{\mathbf{opt}}(l)$ and $h_{\mathbf{sup}}(a',l)$, respectively. These CATEs are not defined with respect to the observed distribution and thus will only be known if they can be uniquely expressed as a functional of the observed distribution, i.e., if they are point identified. For example, when there is no unmeasured confounding the $l$-CATE for $l \in \mathcal{L}$ is identified by $\mathbb{E}(Y | A = 1, L = l) - \mathbb{E}(Y | A = 0, L = l)$. Optimal regimes are also defined with respect to the joint distribution of $(Y^{a = 1}, Y^{a = 0}, A, L)$ and so are generally unknown \textit{a priori} to an investigator. Thus, assumptions must be imposed for the observed data to be linked to these regimes.  However, as elements in a non-euclidean space, we cannot simply extend conventional definitions of identification that apply to real-valued estimands. Therefore, we propose the following definition of identification of a regime: 

\begin{definition}[Identification of a regime]
\label{def: point id regimes}
    Let $\mathcal{M}$ be a set of joint distributions $P^F$ of $(Y^{a=1}, Y^{a=0}, Y,A,L)$. Let $P\equiv P(P^F)$ be the observed margin of $P^F$ and let $\mathbf{M} \coloneqq \{P(P^F): P^F \in \mathcal{M}\}$ be the set of observed margins for the distributions in $\mathcal{M}$. Let $\mathcal{M}_P \equiv \{P^F \in \mathcal{M}: P(P^F)=P\}$ be the set of distributions $P^F$ with the common observed margin $P$. Let $\Pi$ be the set of functions mapping elements in $\mathcal{L} \times \{0,1\}$ to $\{0,1\}$, and consider a fixed function $f:\mathbf{M} \to \Pi$. Let $g_{P^F}\in\Pi$ be a regime that depends on the joint distribution $P^F$. 

    A regime $g_{P^F}$ is \textit{partially locally} identified at $(a',l)$ for the observed distribution $P$ by $f$ if, for each $P^F\in\mathcal{M}_P$, $g_{P^F}(a',l) = f(P)(a',l)$. In short, we say that the regime is $(a',l)$-identified.  
    
    
    Furthermore, a regime is \textit{fully locally} identified at $P$ if it is $(a',l)$-identified at $P$ for all $(a',l) \in \mathcal{L} \times \{0,1\}$. We say that a regime is locally unidentified at $P$ if it is not $(a',l)$-identified for any $(a',l) \in \mathcal{L} \times \{0,1\}$. 

    A regime is \emph{globally} $(a',l)$-identified when it is \emph{locally} $(a',l)$-identified for all laws $P_F$.
    \label{def: Identified regime}
\end{definition}
Similarly, we say the the $l$-optimal regime is identified at $l$, or $l$-identified, for the observed distribution $P$ by $f$ if, for each $P^F\in\mathcal{M}_P$, $g_{\textbf{opt}}(l) = f(P)(l)$.

To fix ideas about Definition \ref{def: Identified regime}, we will discuss some examples. First, consider a trivial case: a static regime $g_{P_F}$ where treatment $a$ is assigned to every individual. This regime is globally fully identified by the function $f: (a',l) \mapsto a$ because $g_{P_F}$ is \textit{defined} to be equal to this $f$ for all laws $P_F$. 

As a second example, consider an $f$ taken as an indicator that an identification formula of the $(L,A)$-CATE is greater than $0$. If all of the $(L,A)$-CATEs are identified by the corresponding identification formula, then the superoptimal regime is globally fully identified by $f$. When only bounds of the $(L,A)$-CATEs are available, then there in general will not exist an $f$ that globally fully identifies the superoptimal regime, although locally full identification may be possible: for example, a law $P$ for which all possible $(l,a)$-CATEs are either above or below 0, for each $(l,a)$. However, when there is no unmeasured confounding, the optimal regime will be equal to the superoptimal regime \citep{StensrudSarvet2022} and the $L$-CATEs will be point identified, which implies that the optimal and superoptimal regimes are globally fully identified. Assumptions, such as the \emph{no unmeasured confounding} assumption, can allow for global identification of a regime, but in practice we only get data from one observed law $P$, so local identification is relevant;  investigators are interested in local identification under the model $P$ that generated the observed data. We will henceforth omit the term \textit{local} and \textit{global} when the context is obvious.

If the $(l,a')$-CATE is point identified, we say that $g_{\textbf{sup}}$ is $(a',l)$-identified or, equivalently, that $g_{\textbf{sup}}(a',l)$ is identified. When $g_{\textbf{sup}}$ is $(a',l)$-identified for all $l$, we will write that $g_{\textbf{sup}}$ is $(a',L)$-identified. For example, suppose that the average effect of treatment in the treated (ATT) is point identified but not the average effect of treatment in the untreated (ATU) \citep{tchetgen2013alternative, huber_sharp_2017}. Then the superoptimal regime is globally $(1,L)$-identified but may be merely locally $(0,L)$-identified, locally $(0,l)$-identified for some $l$, or not identified at all. Such settings are not a technical curiosity; for example, some classical approaches imply point identification of the ATT without identification of the ATU \citep{heckman_matching_1997,abadie_semiparametric_2005,geneletti2011att,tchetgen2013alternative, sarvet2020graphical}.





We emphasize that full identification of a regime will not imply full identification of its value function. In the next section, we will consider conditions that allow (partial) identification of $(L,A)$-optimal regimes and their value functions $\mathbb{E}[Y^{g_{\text{\textbf{sup}}}}]$. We specifically study settings where partial identification of the conditional value functions implies partial identification of $(L,A)$-optimal regimes and their value functions. Indeed, we show that even if the $l$-optimal regime for a given $l \in \mathcal{L}$ is unidentified, we might identify the $(l,0)$-optimal regime, the $(l,1)$-optimal regime or both.

\subsection{Assumptions and identification results}
\label{sec:First_As}
Unlike conventional point identification strategies for conditional average treatment effects, we will not invoke the usual exchangeability assumption, $Y^a \independent A \mid L$; informally, we consider settings where there might be unmeasured confounding. For our initial results, we only rely on the conventional consistency and positivity assumptions, which are routinely invoked in studies of causal effects, see e.g. \citet{hernan2018causal}.
\begin{assumption}[Consistency]
 $Y = Y^a$ when $A=a$ w.p.1 for all $a \in \{0,1\}$.
 \label{ass: cconsistency}
\end{assumption}
\begin{assumption}[Positivity]
$P(A = a \vert L) > 0$ w.p.1 for all $a \in \{0,1\}$.
 \label{ass: positivity}
\end{assumption}

Assumptions \ref{ass: cconsistency} and \ref{ass: positivity}  are sufficient to derive relations between different types of optimal regimes, using the following lemma that has appeared in previous work on treatment effects on the treated \citep{robins2007causal,geneletti2011att,bareinboim2015bandits,huber_sharp_2017,dawid2022can, StensrudSarvet2022}.
\begin{lemma}
Under Assumptions \ref{ass: cconsistency} and \ref{ass: positivity}, 
\begin{align}
 \mathbb{E}(Y^{a} \mid A = a', L=l)
 & =  \begin{cases}
                    \mathbb{E}(Y \mid A = a', L=l),   & \text{if } a = a' , \\
                    \frac{\mathbb{E}(Y^{a} \mid L=l) - \mathbb{E}(Y \mid A = a, L=l)P(A =a \mid L = l)   }{P(A =a' \mid L = l)},  & \text{if }  a \neq a'. 
                \end{cases} \label{eq: lemma}
\end{align}
\label{lemmma1}
\end{lemma}

Lemma \ref{lemmma1} motivates bounds on the conditional value function $\mathbb{E}(Y^{a} \mid A = a', L=l)$ defined as functions of bounds on $\mathbb{E}(Y^a |L = l)$. That is, if $a \neq a'$,  upper and lower bounds on $\mathbb{E}(Y^a | L = l)$ can be plugged into Equation \ref{eq: lemma} to find upper and lower bounds on $\mathbb{E}(Y^a | A = a', L = l)$, respectively. However, we also aim to make statements about sharpness, that is, whether valid bounds can be narrowed:

\begin{definition}[Sharp and valid bounds]

Let $\theta:\mathcal{M} \to \mathbb{R}$ be some finite real-valued parameter. Consider two functions, $f_1:\mathbf{M}\to \mathbb{R}$ and $f_2:\mathbf{M}\to \mathbb{R}$ such that the following property holds for all $P\in \mathbf{M}$,
\begin{itemize}
    \item [] $f_1(P) = \inf\limits_{\mathcal{M}_P}\theta(P^F)$, and  $f_2(P) = \sup\limits_{\mathcal{M}_P}\theta(P^F)$. 
\end{itemize}

Then we say that, $f_1$ and $f_2$ identify \textit{sharp} lower and upper bounds on $\theta$, respectively.  Consider two additional functions, $f'_1$ and $f'_2$ such that for all $P\in \mathbf{M}$, $f'_1(P) \leq f_1(P)$ and $f'_2(P) \geq f_2(P)$. Then we say that $f'_1$ and $f'_2$ identify \textit{valid} bounds on $\theta$ (with respect to $\mathcal{M}$): for all $P\in\mathbf{M}$, $\big(f_1(P), f_2(P)\big) \subseteq  \big(f'_1(P), f'_2(P)\big)$.

\end{definition}

Intervals based on sharp bounds are contained within intervals based on valid bounds. Clearly, sharp bounds are desirable. When convenient, we will suppress notation indicating the dependence on $P$ of the functionals for the bounds. Henceforth we take $\mathbf{L}_a(l)$ and $\mathbf{U}_a(l)$ to be valid lower and upper bounds, respectively, of $\mathbb{E}(Y^a | L = l)$ with width $\phi(a,l):= \mathbf{U}_a(l) - \mathbf{L}_a(l)$, and we take $\mathbf{U}(l) := \mathbf{U}_1(l) - \mathbf{L}_0(l)$ and $\mathbf{L}(l) := \mathbf{L}_1(l) - \mathbf{U}_0(l)$ to be valid upper and lower bounds on the $l$-CATE, respectively. We will always use bold font for bounds; $\mathbf{L}$ and $\mathbf{U}$ should not be confused with $L$ and $U$, or the domains $\mathcal{L}$ and $\mathcal{U}$.

The next proposition gives algebraic relationships between the bounds on $\mathbb{E}(Y^a | A = a', L = l)$, $\mathbb{E}(Y^a |L = l)$, and $ \mathbb{E}(Y^{a = 1} - Y^{a = 0} \mid A = a', L=l)$. We abuse our notation of bounds of the $L$-CATEs; $\mathbf{L}$ and $\mathbf{U}$ are the lower and upper bounds of the $L$-CATE if written as functions of $L$, and are lower and upper bounds of the $(L,A)$-CATE if written as a function of $L$ and $A$. Specifically, we define the functions $\mathbf{L}(l,a')$ and $\mathbf{U}(l,a')$ as
\begin{align*}
     \mathbf{L}(l,a') &\coloneqq (1-2a')\frac{\mathbf{B}_{1-a'}(l) - \mathbb{E}(Y \vert L = l)}{P(A = a'\vert L = l)}, \text{ and }
    \mathbf{U}(l,a')  \coloneqq (1-2a')\frac{\mathbf{B}'_{1 - a'}(l) - \mathbb{E}(Y \vert L = l) }{P(A = a'\vert L = l)}, 
\end{align*}
where $\mathbf{B}_{1-a'}(l) = a'\mathbf{U}_{0}(l) + (1-a')\mathbf{L}_{1}(l)$ and $\mathbf{B}'_{1 - a'}(l) = a'\mathbf{L}_{0}(l) + (1-a')\mathbf{U}_{1}(l)$ are valid lower and upper bounds of $\mathbb{E}(Y^{1-a'} | L = l)$.

\begin{proposition}[Induced bounds]
Suppose Assumptions \ref{ass: cconsistency} and \ref{ass: positivity} hold. Then, $\mathbf{L}(l,a')$ and $\mathbf{U}(l,a')$ are valid bounds on the $(l, a')$-CATE with width $\rho(a',l) := \phi(1-a',l)/P(A = a' | L = l)$. Furthermore, they are \textit{sharp} bounds if and only if the bounds $\mathbf{L}_a(l)$ and $\mathbf{U}_a(l)$ are sharp for $\mathbb{E}(Y^a | L = l)$.

\label{Equiv_PartId}
\end{proposition}

A simple corollary is that, under the same assumptions, sharp bounds on the $(l,a)$-CATE imply sharp bounds on the value functions $\mathbb{E}(Y^a | A = a', L = l)$, $a \neq a'$, see online Appendix \ref{app: Sharpness} for details and a proof of Proposition \ref{Equiv_PartId}. This result on sharpness is useful because previous work on sharp bounds of the value function $\mathbb{E}(Y^a | L = l)$ in different settings, see, e.g., \citet{BalkePearl1997,swanson2018partial,kallus2018interval,Yadlowsky2022}, can be used to derive bounds on $(L,A)$-optimal regimes and value functions. \citet{huber_sharp_2017} also suggested that a result similar to Proposition \ref{Equiv_PartId} gives sharp bounds on average treatment effects of the treated in an IV setting.

The next corollary clarifies that the superoptimal regime $g_{\textbf{sup}}$ is $(a,l)$-identified precisely when the sign of a so-called population intervention effect is identified \citep{hubbard2008population}.      

\begin{corollary}
Under Assumptions \ref{ass: cconsistency} and \ref{ass: positivity}, $g_{\textbf{sup}}(a',l)$ is identified if and only if \\
$\sign \{ \mathbb{E}(Y | L = l) - \mathbb{E}(Y^{1-a'} | L = l) \}$ is identified.
\label{Cor:diff}
\end{corollary}

Thus, we can find the sign of the $(l,a')$-CATE by merely studying a conditional expectation of observed data, $\mathbb{E}(Y | L = l)$, and a conventional conditional value function, $\mathbb{E}(Y^{1-a'} | L = l)$. In comparison, $g_{\textbf{opt}}$ is $l$-identified if and only if the sign of the $l$-CATE is identified, that is, $\sign \{ \mathbb{E}(Y^{a'} | L = l) - \mathbb{E}(Y^{1-a'} | L = l) \}$, which is different from the population intervention effect. Note that $l$-identification of $g_{\textbf{opt}}$ requires $l$-identification of the sign of a contrast involving two counterfactual parameters, whereas $l$-identification of $g_{\textbf{sup}}$ involves only one of those counterfactual parameters. We leverage this to derive improved results for super-optimal regimes. The following proposition states relations between bounds and signs of $l$-CATEs and $(l,a)$-CATEs. 
\begin{proposition} 
Under Assumptions \ref{ass: cconsistency} and \ref{ass: positivity}, the $(l, 1-a')$-CATE $h_{\mathbf{sup}}(1-a',l)$ verifies
\begin{equation*}
    \frac{\mathbf{L}(l) - \mathbf{U}(a',l)P(A = a' | L = l)}{P(A = 1 - a' | L = l)} \leq h_{\mathbf{sup}}(1-a',l) \leq \frac{\mathbf{U}(l) - \mathbf{L}(a',l)P(A = a' | L = l)}{P(A = 1 - a' | L = l)}.
\end{equation*}
If $g_{\textbf{opt}}(l) \neq g_{\textbf{sup}}(a',l)$ for some $a',l$, then $g_{\textbf{sup}}(1-a',l) = g_{\textbf{opt}}(l)$.
\label{Sign_prop}
\end{proposition}
It follows from Proposition \ref{Sign_prop} that having bounds on the $l$-CATE and one of the $(l,a')$-CATEs is sufficient to derive bounds of the $(l, 1-a')$-CATE. We give an example of the sign properties of Proposition \ref{Sign_prop} in online Appendix \ref{app: Reversal}. We can use the width of the intervals given by the bounds to quantify the information we gain by using the natural treatment value to make decisions; informally, for all $l \in \mathcal{L}$, we distribute the width of the bounds on the $l$-CATE to the width of the $(l,a')$-CATEs for $a' \in \{0,1\}$. This notion is formalized in the next proposition.

\begin{proposition}
Let $\omega(l) := \mathbf{U}(l) - \mathbf{L}(l)$. Then, under Assumptions \ref{ass: cconsistency} and \ref{ass: positivity}, $$\omega(l) = P(A = 0| L = l)\rho(0,l) + P(A = 1 | L = l)\rho(1,l),$$ where $\rho(a',l)$ is the width of the induced bounds of the $(l,a')$-CATE.
\label{width_prop}
\end{proposition}
The width of the $l$-CATE is a convex combination of the widths of the $(l,a')$-CATEs for $a' \in \{0,1\}$ and therefore it is contained within $[\min(\rho(0,l), \rho(1,l)), \max(\rho(0,l),\rho(1,l))]$. Thus for some $a' \in \{0,1\}$, $\rho(a',l) \leq \omega(l) \leq \rho(1-a',l)$.

Proposition \ref{width_prop} clarifies that the width of the intervals for one $(l,a')$-CATE, say $\mathbb{E}(Y^{a = 1} - Y^{a = 0} | A = a', L = l)$, is larger than, or equal to, the width of the interval of the $l$-CATE, $\mathbb{E}(Y^{a = 1} - Y^{a = 0} | L = l)$. However, the interval of the complementary $(l,1 - a')$-CATE, $\mathbb{E}(Y^{a = 1} - Y^{a = 0} | A = 1-a', L = l)$, is smaller than, or equal to, the width of the $l$-CATE interval. 

For a given $l$, there exist settings where  $g_{\textbf{sup}}$ is $(l,a')$-identified for all $a' \in \{0,1\}$, even if $g_{\textbf{opt}}$ is not $l$-identified; that is, by using the natural treatment value, we can sometimes identify the superoptimal regime, even if the conventional optimal regime is only partially identified, or not identified at all, see online Appendix \ref{app: Reversal} for details. The next corollary further clarifies the relationship between $g_{\textbf{sup}}(a',l)$ and $g_{\textbf{sup}}(1-a',l)$ when the $l$-optimal regime is not identified, but the $(l,A)$-optimal regime is identified.

\begin{figure}
    \centering
    \includegraphics[scale = 0.3]{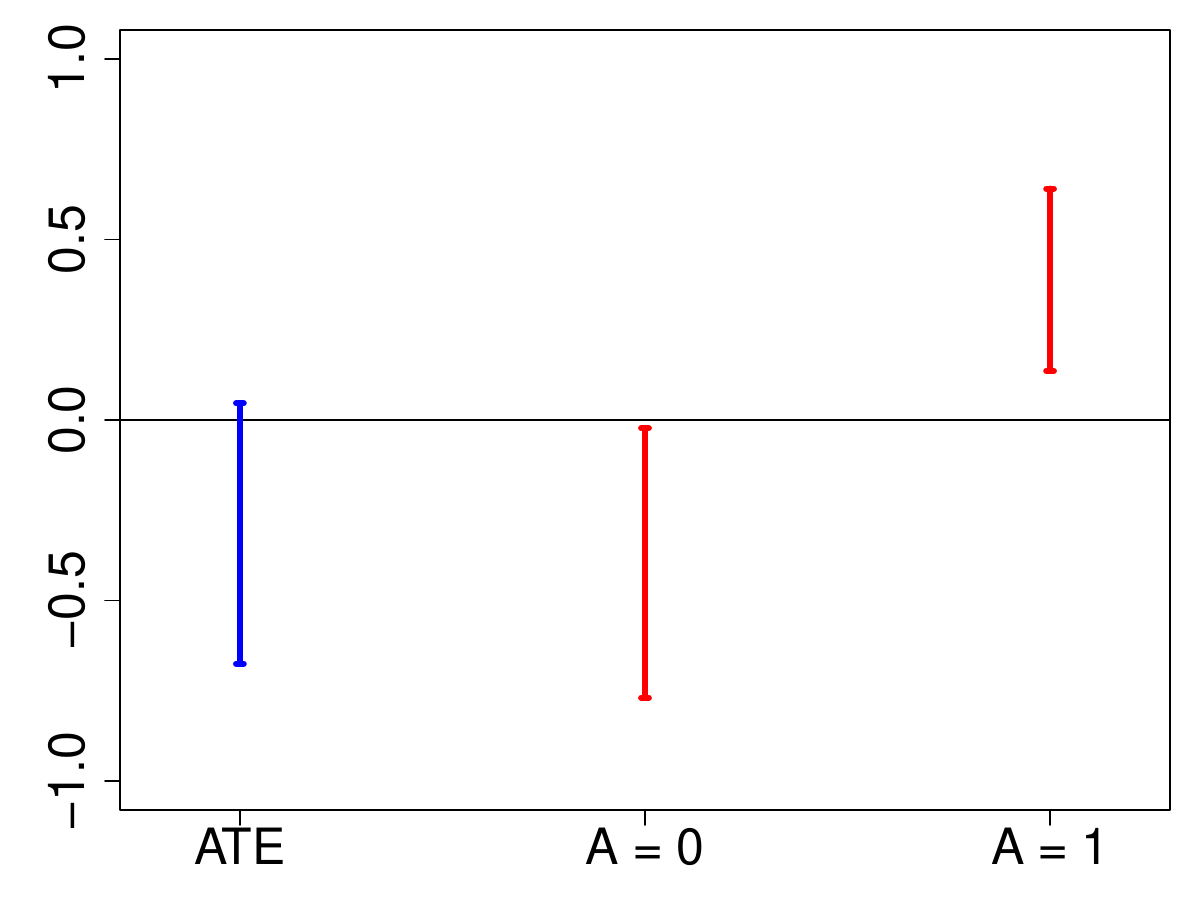}
    \caption{Example of bounds from a distribution that satisfies the statement of Corollary \ref{noCATE_Id}. See online Appendix \ref{app: Reversal} for more details on the distribution.}
    \label{fig:Reversal main}
\end{figure}

\begin{corollary}
Suppose that the $l$-optimal regime is not locally identified under Assumptions \ref{ass: cconsistency} and \ref{ass: positivity}, but the $(l,A)$-optimal regime is locally identified for $P$. Then $g_{\textbf{sup}}(a',l) = 1 - g_{\textbf{sup}}(1-a',l)$. 

\label{noCATE_Id}
\end{corollary}
  Thus, under the conditions of Corollary \ref{noCATE_Id}, the $(l,a')$- and $(l,1-a')$-CATEs have opposite signs. Even in simple graphical models, there exist laws where the $l$-optimal regime is not identified but the $(l,A)$-optimal regime is identified, that is, the optimal regime is unidentified, but the superoptimal regime is $(a',l)$-identified for all $a'\in \{0,1\}$.  A numerical example is shown in Figure \ref{fig:Reversal main}, describing a simple IV setting, where for simplicity we let $\mathcal{L}\equiv\emptyset$; while the ATE covers the null effect, the bounds of both $A$-CATEs are informative, in the sense that the superoptimal regime is identified for all $a' \in \{0,1\}$, but the optimal regime is not. See details about the example in online Appendix \ref{app: Reversal}.

The next corollary further clarifies that the $(l,a')$-optimal regime for some $a'$ is always identified when the $l$-optimal regime is identified. 

\begin{corollary}
    When the $l$-optimal regime is identified under Assumptions \ref{ass: cconsistency} and \ref{ass: positivity}, then the $(l,a')$-optimal regime is identified for some $a'$. 
    \label{cor: l-opt identified}
\end{corollary}
In Appendix \ref{app: BP liability judgment} we illustrate the practical meaning of Corollary \ref{cor: l-opt identified} in an example from \citet{BalkePearl1997} on a fictional liability judgment. Furthermore, the bounds of the $l$-CATE and $(l,a')$-CATEs are only identical in very restrictive settings: 
\begin{corollary}
Under Assumptions \ref{ass: cconsistency} and \ref{ass: positivity}, for any $a \in \{0,1\}$, $\mathbf{U}(l) = \mathbf{U}(l,a)$ and $\mathbf{L}(l) = \mathbf{L}(l,a)$ if and only if $\mathbf{U}(l) = \mathbf{U}(l,a) = \mathbf{U}(l, 1-a)$ and $\mathbf{L}(l) = \mathbf{L}(l,a) = \mathbf{L}(l, 1- a)$.
\label{Bound_equ}
\end{corollary}


Based on the results in this section, we have a simple algorithm for creating (sharp) bounds of $(l,a')$-CATEs and identifying $(l,a')$-optimal regimes, given that (sharp) bounds of $\mathbb{E}[Y^a | L = l]$ are available;  we use Lemma \ref{lemmma1} to compute induced bounds of the $(l,a')$-CATEs using the bounds of $\mathbb{E}[Y^a | L = l]$ and then identify the $(l,a')$-optimal regime when the sign of the $(l,a')$-CATE is identified, see Algorithm \ref{Algo: (L,A)-CATE bounds and superopt} in online Appendix \ref{app: superopt algo}.

\subsection{Decision criteria ensuring globally identified regimes}
\label{sec: Decision-making criteria}
The bounds on $l$-CATEs and $(l,a')$-CATEs might cover null effects for some laws $P$, such that the $l$-optimal and $(l,a')$-optimal regimes are not globally fully identified. However, the decision maker can still seek to maximize the expected outcome using decision criteria, which are globally fully identified by design when utility parameters are partially identified, see for example \citet{manski_identification_2000,manski_treatment_2002,stoye_essays_2005}. In particular, \citet{Cui2021Individualized} summarized decision criteria for partially identified CATEs and we can adapt these criteria to the $(L,A)$-optimal regimes. As an illustration, we consider four canonical examples of decision criteria given covariates $L$ (Table \ref{fig:Decision-making criteria}). Some of the decision criteria are similar to known decision-making models, for example, the minimax regret criterion  (opportunist) resembles the \citet{Roy1951} model in economics. Using the terminology of \citet{Roy1951} and \citet{eisenhauer2015}, the opportunist compares the maximum benefit of choosing $a = 1$ with the maximum cost that can be incurred when choosing $a = 1$ compared to the alternative $a = 0$. Analogously, the opportunist will choose the policy that maximizes their expectation when assigning a uniform prior on the bounds of the CATE, in accordance with the \citet{Roy1951} model. \citet{Cui2021Individualized} showed that many decision criteria that are functions of $L$ are optimal with respect to corresponding risk functions. We show this also holds for decision criteria that are functions of $A$ and $L$, see online Appendix \ref{app: Decision-making criteria}.

Here we also define two different criteria that incorporate the natural treatment value. 
\begin{definition}[Superoptimal conventionality criterion]
     The superoptimal conventionality criterion is defined as
     \begin{align*}
         g_{A - \text{sup}}(a',l) := (1-2a')I((1-a')\mathbf{L}(a',l) - a'\mathbf{U}(a',l) > 0) + a'.
     \end{align*}
     \label{def: superopt healthcare with obs baseline}
 \end{definition}
\begin{definition}[Conventionality criterion]
     The optimal conventionality criterion is given by
     \begin{align*}
         g_{A - \text{opt}}(a',l) &:= a'(1-I(\mathbf{L}_0(l) > \mathbb{E}[Y | L = l])I(\mathbf{L}_1(l) \leq \mathbf{U}_0(l)))\\
         &+ (1-a')I(\mathbf{L}_1(l) > \max(\mathbf{U}_0(l), \mathbb{E}[Y | L = l])).
     \end{align*}
     \label{def: opt healthcare with obs baseline}
\end{definition}

The superoptimal conventionality criterion ensures that $g_{A - \text{sup}}(a',l)$ assigns the observed natural treatment value $a'$, unless the bounds of the $(l,a')$-CATE identify $g_{\textbf{sup}}(a',l) = 1-a'$. Similarly, the optimal conventionality criterion uses the bounds of the $L$-CATE.

The optimal conventionality criterion gives a regime that coincides with the observed regime for every $l$, unless the bounds show that giving either treatment $a = 1$ or $a = 0$ outperforms the observed regime. 

The conventionality criteria are similar to the so-called healthcare decision criterion \citep{Cui2021Individualized}, except that the observed regime is the baseline regime. This observation motivates the following results:
\begin{proposition}
    The conventionality criteria ensure that the observed regime is outperformed, $$\mathbb{E}[Y^{g_{A - \text{sup}}}]\geq \mathbb{E}[Y^{g_{A - \text{opt}}}] \geq \mathbb{E}[Y].$$
    \label{prop: superopt healthcare with obs base}
\end{proposition}
\begin{corollary}
    For any $l \in \mathcal{L}$, 
    \begin{align*}
        \mathbb{E}[Y | L = l] \leq \mathbf{L}_{g_{A-\text{opt}}(0,l)}(l) = \mathbf{L}_{g_{A-\text{opt}}(1,l)}(l) \leq \mathbb{E}[Y^{g_{A - \text{sup}}}| L = l].
    \end{align*}
    \label{cor: superopt healthcare with obs base}
\end{corollary}

We compare the conventionality criterion to other decision criteria in our data analysis in Section \ref{sec: ICU example}.

\begin{table}
    \centering
    \resizebox{\hsize}{!}{\begin{tabular}{ |p{3.5cm}||p{7.5cm}|p{7cm}|}
 \hline
 Decision criterion & Definition & Regime\\
 \hline
 Maximax utility (Optimist)   & $\max_{g}\max_{P^F \in \mathcal{M}_P} \mathbb{E}(Y^g | L = l)$ & 
 \resizebox{0.45\hsize}{!}{$g(l) = \begin{cases}
        1 & \mathbf{U}_1(l) > \mathbf{U}_0(l)\\
        0 & \mathbf{U}_1(l) < \mathbf{U}_0(l)
    \end{cases}$}\\
 Maximin utility (Pessimist) & $\max_{g}\min_{P^F \in \mathcal{M}_P} \mathbb{E}(Y^g | L = l)$  & \resizebox{0.45\hsize}{!}{$g(l) = \begin{cases}
        1 & \mathbf{L}_1(l) > \mathbf{L}_0(l)\\
        0 & \mathbf{L}_1(l) < \mathbf{L}_0(l)
    \end{cases}$} \\
 Minimax regret (Opportunist) & $\min_{g} \max_{P^F \in \mathcal{M}_P}( \mathbb{E}(Y^{g^*}) - \mathbb{E}(Y^{g}) | L = l)$ & \resizebox{\hsize}{!}{$g(l) = \begin{cases}
        1 & \mathbf{L}(l) > 0 \text{ or } \mathbf{L}(l) < 0 < \mathbf{U}(l), |\mathbf{U}(l)| > |\mathbf{L}(l)| \\
        0 & \mathbf{U}(l) < 0 \text{ or } \mathbf{L}(l) < 0 < \mathbf{U}(l), |\mathbf{U}(l)| < | \mathbf{L}(l)|
    \end{cases}$} \\
 Healthcare ($a = 0$ baseline) & $\max_{g}   \mathbb{E} \{ \mathbb{E}(Y^0 \mid L) + \mathbf{L}(L) g(L)  | L = l \}$ & \resizebox{0.38\hsize}{!}{$g(l) = \begin{cases}
        1 & \mathbf{L}(l) > 0 \\
        0 & \mathbf{L}(l) < 0
    \end{cases}$}\\
    \hline
\end{tabular}}
    \caption{Decision criteria from \citet{Cui2021Individualized}.}
    \label{fig:Decision-making criteria}
\end{table}

\subsection{Superoptimal regimes in practice}


A potential objection to using superoptimal regimes is that the natural treatment might change when individuals know that it will be used to make a decision, a type of strategic action \citep{munro_treatment_2023, kido_incorporating_2023}. Another concern is that future decision makers, who will implement the superoptimal regimes, might differ from those who made decisions in the observed data \citep{ida_choosing_2022}. For the guarantees in Section \ref{sec:Identification} to be valid, the observed data must be representative of the future decision setting in a certain sense; that is, the results on the natural treatment value in Section \ref{sec:Identification} require a particular stability of the probabilistic structure of treatment, covariates, and outcome in future settings. It is sufficient to assume that the distributions of $Y^a$ and $A$ conditional on $L$, $f_{Y^a | L}$ and $f_{A|L}$, are preserved \citep{StensrudSarvet2022}. Requiring some stability of the probabilistic structure is not unique to $(L,A)$-optimal regimes and is also a feature of $L$-optimal regimes \citep{bareinboim2013general, dahabreh2019generalizing}, although assumptions about the stability of $f_{A|L}$ would not typically be necessary.

To mitigate these concerns, \citet{StensrudSarvet2022} proposed a way of using $(L,A)$-optimal regimes that keeps the decision maker in control of the final decision while informing them of the two $(L,A)$-optimal regimes, allowing them to keep their natural treatment value private. For each covariate value $l$, \citet{StensrudSarvet2022} recommended that  the $(l,a')$-optimal regimes for $a'\in\{0,1\}$ are given to the decision maker, who then makes an action based on this extra information, see online Appendix \ref{app: Use of superopt regime} for more details. In this procedure, the decision-maker need not reveal their natural treatment intention, nor disclose whether the algorithm's recommendation was followed. However, any (human-induced) deviation from the algorithm implies that the $(L,A)$-optimal regime may not outperform the $L$-optimal regime.  

Even if it is impractical or unreasonable to use the natural treatment value in future settings, the $(L,A)$-optimal regime is still a relevant parameter to study in observational data; this regime can reveal if decision makers in the observed data were using relevant information on unmeasured covariates and if they were outperforming the optimal regime \citep{sarvet_perspective_2023, mueller_perspective_2023, sarvet2024rejoinder}. In particular, the $(L,A)$-optimal regime can detect situations where decision makers made undesirable actions. One example is iatrogenic harm; \citet{StensrudSarvet2022} locate an infamous case in the history of medicine -- Ignaz Semmelweis's discovery of inappropriate sanitary conditions -- as one setting where $(L,A)$-optimal regimes may have been useful as a surveillance technology. 

\section{Case study: A Marginal Sensitivity Model}
\label{sec: ATE case study}

The results derived in Section \ref{sec:Identification} can be applied to any setting where valid bounds on the value function $\mathbb{E}(Y^a \mid L = l)$ are computed from the joint distribution  of $(Y,A,L)$. Hence, these results are applicable to any setting where investigators aim to study $l$-optimal regimes in the presence of unmeasured confounding. As a concrete application of these results, we first consider the Marginal Sensitivity Model \citep{Tan2006,zhao2018sensitivity, kallus2018interval}, which extends Rosenbaum's sensitivity model \citep{Rosenbaum2002}. Let $\Gamma \geq 1$ be a user-specified constant, and consider the following assumption.

\begin{assumption}[Marginal Sensitivity Model \citep{Tan2012,zhao2018sensitivity,kallus2018interval}]
\label{ass: marginal sensitivity model}
For any $a \in \{0,1\}, l \in \mathcal{L}, y \in \mathcal{Y}$,
\begin{equation}
    \frac{1}{\Gamma} \leq \frac{(1 - P(A = a | L = l))P(A = a | L = l, Y^a = y)}{P(A = a | L = l)(1 - P(A = a | L = l, Y^a = y))} \leq \Gamma.
    \label{eq: marginal sensitivity model assumption}
\end{equation}
\end{assumption}
When there is no unmeasured confounding, $P(A = a | L = l, Y^a = y) = P(A = a | L = l)$ and Equation \eqref{eq: marginal sensitivity model assumption} holds with $\Gamma = 1$. Larger values of $\Gamma$ allow for stronger unmeasured confounding. 

Assumption \ref{ass: marginal sensitivity model} holds if there exists $\epsilon > 0$ such that for all $a \in \{0,1\}, l \in \mathcal{L}, y \in \mathcal{Y}$, we have that $\epsilon \leq P(A = a | L = l, Y^a = y) \leq 1 - \epsilon$ and $\epsilon \leq P(A = a | L = l) \leq 1 - \epsilon$ (choose $\Gamma = (1-\epsilon)^2/\epsilon^2$).

Assumption \ref{ass: marginal sensitivity model} gives an expression for bounds of $\mathbb{E}(Y^a | L = l)$ \citep{kallus2018interval}, which implies bounds on the CATEs. Furthermore, these bounds are sharp under regularity conditions \citep{kallus2018interval}[Theorem 1]. Now, using Proposition \ref{Equiv_PartId}, we deduce that the induced bounds on the $(l,a')$-CATEs and their value functions are sharp under the same conditions. To illustrate the implications of these results, we analyzed the simulated data example from Section 5 of \citet{kallus2018interval}: 

\begin{example}[\citet{kallus2018interval}]
\label{ex: Kallus sensitivity}
Let $U \sim \text{Bern}(0.5)$ be a binary unobserved common cause of treatment $A$ and outcome $Y$, and define the observed covariate $L \sim \text{Unif}[-2,2]$. Let $e(l) := P(A = 1 | L = l)$ be fixed to $e(l) = \sigma(0.75l + 0.5)$, where $\sigma$ is the sigmoid function. Let $\alpha(l, \Gamma) := 1/(\Gamma e(l)) + 1 - 1/\Gamma$ and $\beta(l, \Gamma) = \Gamma/e(l) + 1 - \Gamma$ and the propensity score $e(l,u) := u/\alpha(l, \Gamma^*) + (1-u)/\beta(l, \Gamma^*)$, where $\Gamma^* = \exp(1)$. 

After drawing $A |L, U \sim \text{Bern}(e(L,U))$, we computed bounds using the estimators in Section 5 of \cite{kallus2018interval}, with a logistic regression for the propensity score with $n = 2000$ observations and for $\Gamma = \exp(0.5)$, $\exp(1)$ and $\exp(1.5)$. Under this data-generating mechanism, the $(L,A)$-optimal is different from the $L$-optimal regime, as illustrated in Figure \ref{fig: Kallus CATEs}. Furthermore, the $(L,A)$-optimal regime is fully identified for $\Gamma = \exp(0.5),\exp(1)$ and partially identified for $\Gamma = \exp(1.5)$, even when the $L$-optimal regime is not, see Figure \ref{fig: Kallus CATEs}, in particular, in the case where the sensitivity model is correctly specified with $\Gamma =  \exp(1.0)$.

\begin{figure}
    \centering
    \includegraphics[scale = 0.36]{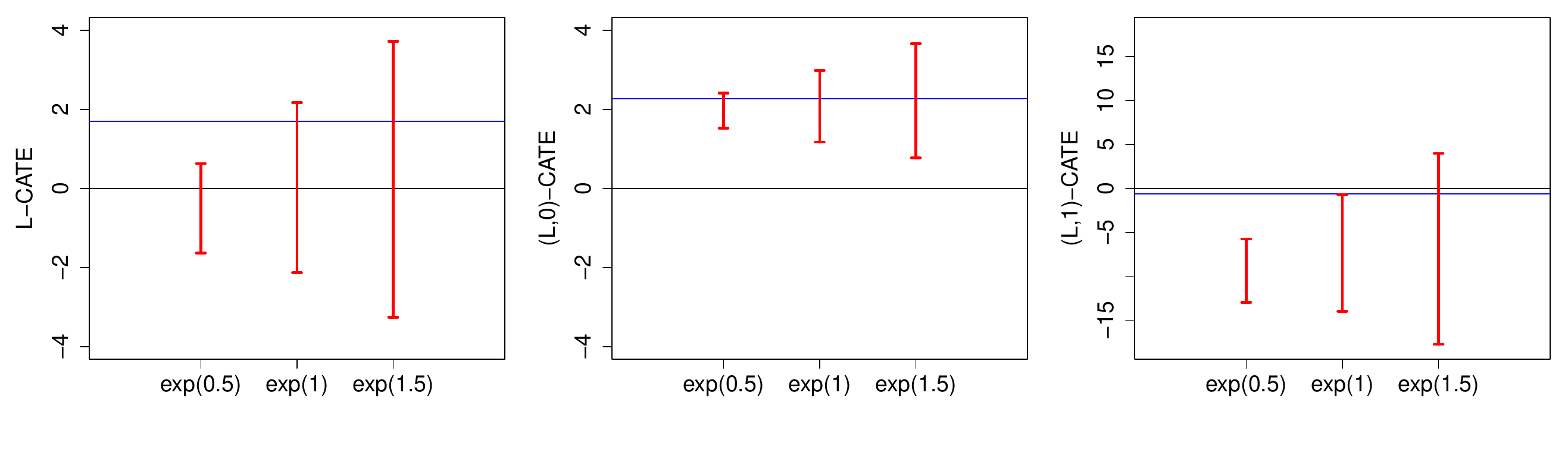}
    \caption{CATE bounds in Example \ref{ex: Kallus sensitivity} for $L = 1.2$, based on the example in \citet{kallus2018interval}. The true values of the CATEs are represented with a blue line.}
    \label{fig: Kallus CATEs}
\end{figure}
\end{example}

\subsection{Sharpness under \citet{Rosenbaum2002}'s sensitivity model}
\citet{Yadlowsky2022} also studied bounds on the $L$-CATE under a sensitivity model, inspired by \citet{Rosenbaum2002}, which invoked an assumption that is related to Assumption \ref{ass: marginal sensitivity model}: 
\begin{assumption}
    \label{ass: Rosenbaum assumption}
    For unmeasured confounders $U$ that can affect $A$ and $Y^a$, and almost all $u, \Tilde{u}$ in the domain of $U$, $L$,
    \begin{equation*}
        \frac{1}{\Gamma} \leq \frac{P(A = 1 | L, U = u)P(A = 0| L, U = \Tilde{u}) }{P(A = 0 | L, U = u)P(A = 1| L, U = \Tilde{u})} \leq \Gamma.
    \end{equation*}
\end{assumption}
Like Assumption \ref{ass: marginal sensitivity model}, it follows that Assumption \ref{ass: Rosenbaum assumption} holds with $\Gamma = 1$  when there is no unmeasured confounding. Furthermore, Assumption \ref{ass: Rosenbaum assumption} holds if there exists $\epsilon > 0$ such that $\epsilon \leq P(A = a | L, U) \leq 1 - \epsilon$ almost surely for all $a \in \{0,1\}$; we can, for example, choose $\Gamma = (1-\epsilon)^2/\epsilon^2$. 

Under Assumption \ref{ass: Rosenbaum assumption}, \citet{Yadlowsky2022}[Section 3.4] derived bounds on $\mathbb{E}(Y^a | L = l)$, which are sharp under regularity conditions and can be estimated at $\sqrt{n}$ rate if the nuisance parameters are estimated at $n^{1/4}$ rate. However, there are no guarantees that the induced bounds for the $l$-CATE are sharp, as acknowledged by \citet{Yadlowsky2022}. Nevertheless, we know that the induced bounds on the $(l,a')$-CATEs and value functions are sharp if and only if the bounds on $\mathbb{E}(Y^a | L = l)$ are sharp, see online Appendix \ref{app: Sharpness} for details. Thus, although bounds on the $l$-CATE under Assumption \ref{ass: Rosenbaum assumption} may not be sharp, we have sharp bounds on the $(l,a')$-CATEs.

\section{Case Study: Instrumental Variables}
\label{sec: Case Study - IV}
Henceforth, we will consider the binary IV setting as another practically relevant case; IVs are often leveraged to (point and partially) identify causal estimands when there is unmeasured confounding. Therefore, let $Y \in \{0,1\}$ be the outcome and $Z \in \{0,1\}$ be an IV.

\begin{figure}
            \centering
            \begin{tikzpicture}
                \tikzset{line width=1.5pt, outer sep=0pt,
                ell/.style={draw,fill=white, inner sep=2pt,
                line width=1.5pt},
                swig vsplit={gap=5pt,
                inner line width right=0.5pt},
                swig hsplit={gap=5pt}
                };
                      \node[name=L,ell,shape=ellipse] at (3,-2){$L$};
                    \node[name=A,shape=swig vsplit] at (6,0) {
                                                      \nodepart{left}{$A^z$}
                                                      \nodepart{right}{$a$} };
                    \node[name=Y, ell, shape=ellipse] at (9,0){$Y^{a}$};
                    \node[name=U, ell, shape=ellipse] at (3,2){$U$};
                    \node[name=Z, shape=swig vsplit] at (3,0){
                                                      \nodepart{left}{$Z$}
                                                      \nodepart{right}{$z$}};
                \begin{scope}[>={Stealth[black]},
                  every node/.style={fill=white,circle},
                  every edge/.style={draw=black,very thick}]
                    \path[->] (U) edge[bend left=10] (Y);
                    \path[->] (U) edge (A);
                    \path[->] (L) edge (A);
                    \path[->] (Z) edge (A);
                    \path[->] (A) edge (Y);
                    \path [->] (L) edge[bend right=10] (Y);
                \end{scope}
            \end{tikzpicture}
            \caption{
            A SWIG with instrumental variable $Z$ and interventions on $Z$ and $A^z$.
            }
                            \label{fig:superoptimal}
\end{figure}
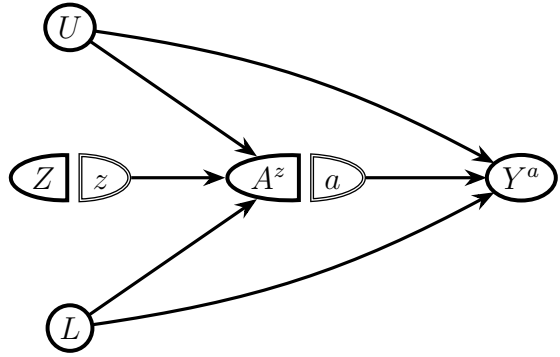

Specifically, consider the Single-World Intervention Graph (SWIG) in Figure \ref{fig:superoptimal}, where we also assume that Assumptions \ref{ass: cconsistency} and \ref{ass: positivity} hold. All variables are observed except $U$. The SWIG in Figure \ref{fig:superoptimal}, when assumed to encode individual level exclusion restrictions, implies the following assumptions:
\begin{assumption}[Unconfounded IV-outcome relation]
For all $a,z$, $Y^{a,z} \independent Z | L$.
    \label{As:Unconf}
\end{assumption}
\begin{assumption}[Exclusion restriction]
For all $a,z$, $Y^{a,z}  = Y^a$.
    \label{As:Excl}
\end{assumption}

In this binary IV setting, when Assumptions \ref{ass: cconsistency} and \ref{ass: positivity} hold, \citet{BalkePearl1997} derived sharp bounds on $\mathbb{E}(Y^a \vert L = l)$ and $\mathbb{E}(Y^{a = 1}- Y^{a = 0} | L = l)$, see online Appendix \ref{app: BP bounds formulas} for details.\footnote{There are also weaker IV conditions under which the Balke-Pearl bounds still hold, see \citet{swanson2018partial} for a review.} These bounds have recently been used to identify the $L$-optimal regime \citep{pu2021estimating, cui2021machine, Cui2021Individualized}. Because the Balke-Pearl bounds of $\mathbb{E}(Y^a \vert L = l)$ are sharp, we can immediately use Proposition \ref{width_prop} to derive sharp bounds for superoptimal regimes. 

\begin{remark}
In an IV setting, \citet{StensrudSarvet2022} showed that using the instrument as input to the decision function can further improve the expected outcomes. These results are also relevant in the setting we consider here, and we give further definitions in online Appendix \ref{appendix:fixed_obs}. See also online Appendix \ref{app: NLSYM analysis conditional on Z} for a data analysis.
\end{remark}

\subsection{Estimation}
\label{sec: Estimation}
To motivate our estimators, we will use that
\begin{align}
    \mathbb{E}(Y^g) &= \sum_{a' \in \{0,1\}, l \in \mathcal{L}} \mathbb{E}(Y^{g(a',l)} | A = a', L = l) P(A = a', L = l),
    \label{eq: regime_bounds}
\end{align}
and that $g(A,L)$ is equal to a constant, $g(a',l)$, conditional on $A = a', L = l$. If $g(a',l) = a'$, then by consistency $\mathbb{E}(Y^{g(a',l)} | A = a', L = l) = \mathbb{E}(Y | A = a', L = l)$. Otherwise if $g(a',l) = 1-a'$, we can bound $\mathbb{E}(Y^{g(a',l)} | A = a', L = l)$ using Lemma \ref{lemmma1}.

\cite{levis2023covariateassisted} recently leveraged baseline covariates ($L$) to narrow the original Balke-Pearl bounds of $\mathbb{E}[Y^a]$ for $a \in \{0,1\}$. In particular, they defined influence function-based estimators of bounds under a margin condition, which were proven to have parametric convergence rates when nuisance functions are estimated with flexible models. We build on the results from \cite{levis2023covariateassisted} to derive estimators of bounds of $(L,A)$-optimal regimes. Here we give an overview of the estimation strategy, and a detailed description is found in online Appendix \ref{app: levis procedure}. 

Firstly, the efficient influence functions of \eqref{eq: regime_bounds} is
\begin{align*}
    \mathbb{IF}[\mathbb{E}(Y^g)] = \sum_{a' \in \{0,1\}, l \in \mathcal{L}} & \mathbb{IF}[\mathbb{E}(Y^{g(a',l)} | A = a', L = l)]P(A = a', L = l)\\
    &+ \mathbb{E}(Y^{g(a',l)} | A = a', L = l)(I(A = a', L = l) - P(A = a', L = l)).
\end{align*}
If $g(a',l) = a'$, then $\mathbb{IF}[\mathbb{E}(Y^{g(a',l)} | A = a', L = l)] = Y - \mathbb{E}(Y | A = a', L = l)$. If $g(a',l) = 1-a'$, we derive the influence function of the bounds of $\mathbb{E}(Y^{g(a',l)} | A = a', L = l)$. Let $\psi_1(a,l)$ be a bound on $\mathbb{E}(Y^a | L = l)$. Define
\begin{equation*}
    \Psi(a',l) := \frac{\psi_1(1-a',l) - \mathbb{E}[Y | A = 1-a', L = l]P(A = 1-a' | L = l)}{P(A = a' | L = l)}.
\end{equation*}

We adapt the results from \citet{levis2023covariateassisted}, to derive the non-parametric efficient influence function for $\Psi(a',l)$: 
\begin{proposition}
 The efficient influence function of $\Psi(a',l)$ is
\begin{align*}
    &  \Psi^{\text{eff}}(a',l) =\\
    & \frac{1}{P(A = a'| L = l)^2}\Bigg(\psi_1^{\text{eff}}(1-a',l)P( A = a' | L = l)\\
    & - \psi_1(1-a',l) (I[A = a'] - P(A = a' | L = l)\\
    & - \bigg(\frac{I[A = 1-a']}{P(A = 1-a' | L = l)}(Y - \mathbb{E}[Y  \mid  A = 1-a', L = l]) P(A = 1-a', L = l)\\
    & + \mathbb{E}[Y  \mid  A = 1-a', L = l](I[ A= 1-a'] - P( A = 1-a' | L = l))\bigg) \cdot P(A = a') \\
    &+ \mathbb{E}[Y  \mid  A = 1-a', L = l] P( A = 1-a' | L = l) (I[A = a'] - P(A = a'| L = l)) \Bigg),
\end{align*}
where $\psi_1^{\text{eff}}(a,l)$ is the \emph{centered} influence function of $\psi_1(a,l)$.
\label{thm:proof eff infl}
\end{proposition}


The influence function $\mathbb{IF}[\mathbb{E}(Y^g)]$ only exists if $\mathbb{E}(Y^g)$ is pathwise differentiable. To ensure pathwise differentiability, we can make a \emph{non-exceptional law} assumption for optimal regimes, that is, the probability of no treatment effect conditional on $L$ (and $A$ for the $(L,A)$-optimal regimes) is zero \citep{robins2004optimal,Luedtke2016}.
 
We use hats to denote plug-in estimators. The influence function in Proposition \ref{thm:proof eff infl} motivates a one-step estimator of $\mathbb{E}[\Psi(a',L)]$,  that is,  
$\hat{\mathbb{E}}[\hat{\Psi}(a',L)] + \mathbb{P}_n(\hat{\Psi}^{\text{eff}}(a',L))$, where we use plug-in estimators of $\mathbb{E}(Y| A = a', L = l)$ and $ P(A = a' | L = l)$ for $a' \in \{0,1\}$ to obtain $\hat{\Psi}(a',l)$ and $\hat{\Psi}^{\text{eff}}(a',l)$.
This estimator requires the model for $\psi_1(a',l)$ to be correctly specified. The assumptions of Theorem \ref{thm: Levis} in online Appendix \ref{app: Levis theorem} ensures correct specification and $\sqrt{n}$-convergence of $\mathbb{E}[\Psi(a',L)]$ if the chosen estimators for nuisance parameters converge fast enough, see Proposition \ref{prop: One-step convergence} in the online Appendix for exact rates and conditions. 

Estimating the $(L,A)$-optimal regime $g_{\mathbf{sup}}$ often requires fewer model assumptions than estimating the value function $\mathbb{E}[Y^{g_{\mathbf{sup}}}]$. It follows from Corollary \ref{Cor:diff} that we only need to identify and estimate the difference $\mathbb{E}(Y| L = l)-\psi_1(a',l)$, which has the following non-parametric influence function:
\begin{corollary}
\label{cor: sign id}
The efficient influence function of $\mathbb{E}(Y| L = l)-\psi_1(a',l)$ in $\mathcal{M}_{\text{np}}$ is $Y - \mathbb{E}(Y | L = l) - \psi_1^{\text{eff}}(a',l)$.
\end{corollary}
Hence, a one-step estimator of $\mathbb{E}(Y) - \mathbb{E}[\psi_1(a',L)]$ is $$\hat{\mathbb{E}}(Y) - \hat{\mathbb{E}}[\hat{\psi}_1(a',L)] + \mathbb{P}_n\left(Y - \hat{\mathbb{E}}(Y) - \hat{\psi}_1^{\text{eff}}(a', L) \right).$$ If $\mathbb{E}(Y)$ is estimated nonparametrically, this estimator requires only the model for $\psi_1(a,l)$ to be correctly specified. Proposition \ref{prop: cor 5 convergence result} in online Appendix \ref{app: One-step estimator convergence} describes convergence rates and conditions for the one-step estimator.


In the following data analyses, we compute estimates $\hat{g}_{\textbf{opt}}$ and $\hat{g}_{\textbf{sup}}$ using the decision criteria in Section \ref{sec: Decision-making criteria} and models for the \citet{BalkePearl1997} bounds described in online Appendix \ref{app : NLSYM models}. Then, we estimate the value functions $\mathbb{E}(Y^{\hat{g}_{\textbf{opt}}})$ and $\mathbb{E}(Y^{\hat{g}_{\textbf{sup}}})$ by using the one-step estimator motivated by Proposition \ref{thm:proof eff infl} analogously to \citet{levis2023covariateassisted}. This procedure gives us $\sqrt{n}$-consistent estimators for the value functions of the estimated regimes under the conditions described in Proposition \ref{prop: One-step convergence}.



\section{Effect of prompt ICU admission on survival}
\label{sec: ICU example}
Following \citet{keele2020stronger} and \citet{StensrudSarvet2022}, we studied the effect of prompt admission to the ICU ($A$) on $7$-day survival ($Y$), using information on age, sex and sequential organ failure assessment (SOFA) score, encoded in the covariate vector $L$. We used data resampled from a cohort study of patients with declining health, who were recommended for assessment for ICU admission in 48 hospitals within the UK National Health Service (NHS) in 2010-2011 \citep{harris2015delay}. By prompt admission, we mean that the patient was admitted within four hours of arriving at the hospital. Following \citet{keele2020stronger}, we used an indicator of ICU bed occupancy being higher than the median as an instrument $Z$.

\begin{figure}
    \centering
    \includegraphics[scale = 0.4]{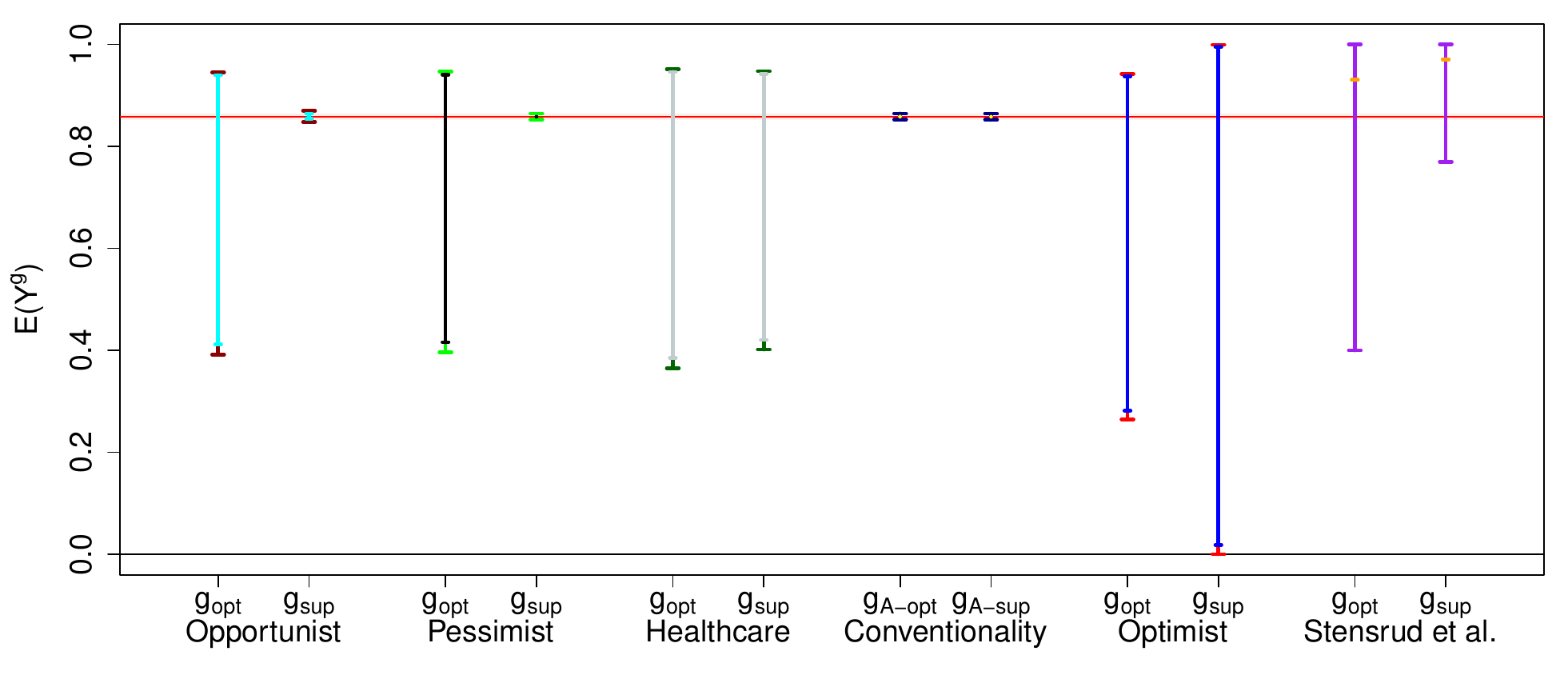}
    \caption{Bounds of value functions in the ICU example.}
    \label{fig:ICU_regime_bounds}
\end{figure}

We computed Balke-Pearl bounds and confidence intervals, using estimators described in Section \ref{sec: Estimation}. Furthermore, we compared the confidence intervals of the bounds with the confidence intervals of point-identified parameters under additional assumptions from \citet{cui2021semiparametric,StensrudSarvet2022}. All the bounds are narrower for regimes that use the natural treatment value except for the optimist bounds. This observation is not surprising because the superoptimal decision rule might often exactly coincide with the natural treatment value, and the outcomes under the natural treatment value are point identified, see Appendix \ref{app: Bound widths} for details. 

In our analysis, the bounds of $\mathbb{E}[Y^{\hat{g}}]$ are sharp. We nevertheless observe that the lower bounds of $\mathbb{E}[Y^{\hat{g}}]$ are smaller than $\mathbb{E}[Y]$ for the optimist and healthcare decision criteria, even if $\mathbb{E}[Y^{g_{\textbf{sup}}}] \geq \mathbb{E}[Y]$. This is not a contradiction: as we estimate the bounds of value functions $\mathbb{E}[Y^{\hat{g}}]$, the estimated bounds of $\mathbb{E}[Y^{\hat{g}}]$ can fluctuate above or below their true value. Furthermore, any decision criterion $g$ that we use in our example does not necessarily verify $\mathbb{E}[Y^{g}] \geq \mathbb{E}[Y]$, as $g(a',l)$ can be different from $g_{\textbf{sup}}(a',l)$ when $\mathbf{L}(l,a') < 0 < \mathbf{U}(l,a')$, which in our case can happen if the \citet{BalkePearl1997} bounds of $\mathbb{E}[Y^{1-a'}|L]$ are wide. A solution for decision makers who want to match or outperform the observed regime, that is, who want a globally fully identified regime $g$ such that $\mathbb{E}[Y^g] \geq \mathbb{E}[Y]$ is to use a decision criterion that uses the observed regime $A$ as a baseline, like the proposed conventionality criteria.

Finally, we observe that the confidence intervals based on bounds are often of similar width or tighter than the confidence intervals for the point estimates of \citet{StensrudSarvet2022}, even though \citet{StensrudSarvet2022} used an additional assumption about effect homogeneity. This can be explained by the difficulty of estimating instrumental variable functionals, whereas estimators for bounds are usually more stable. Moreover, applying the same logic as the decision criteria in Figure \ref{fig:ICU_regime_bounds} for finite samples, some decision makers prefer to implement regimes based on bounds, as the confidence intervals for point identification methods are wide and their validity requires more assumptions.

We have included two additional data analyses in online Appendices \ref{app: Details of NLSYM} and \ref{sec: Influenza regime bounds example}.


\section{Future directions}
Time-varying treatment regimes as functions of time-varying natural treatment values will be of particular interest. Results on optimal time-varying regimes, without the use of natural treatment values, have been studied ubiquitously in settings with no unmeasured confounding, but few results exist when there is unmeasured confounding. Recently, \cite{chen_estimating_2023} considered a time-varying treatment setting with time-varying instruments, but did not leverage the natural treatment value. We also aim to study settings where the treatment is non-binary, and we give a brief description of results for treatments with countable support in online Appendix \ref{app: Non-binary A}.

Furthermore, there exist results that allow us to give confidence intervals for the bounds of $L$-CATEs, for example \citet{ImbensManski20004, JiangDing2018}, but they rely on strong assumptions, see online Appendix \ref{Jiang and Ding procedure}. Deriving results that require less restrictive assumptions is a topic for future investigations, e.g. building on work from \citet{kennedy_towards_2023}.


\if1\blind
{
\subsection*{Acknowledgments}
We would like to thank Dr. Matias Janvin and three anonymous referees for their insightful comments. We are also grateful to the authors of \cite{levis2023covariateassisted} who provided us with the influenza data and code to implement the bounds introduced in their paper. We thank Prof. Luke Keele for providing the ICU data. The authors were supported by the Swiss National Science Foundation, grant $200021\_207436$.

\subsection*{Disclosure statement} The authors have no competing interests to declare.
} \fi
\bibliographystyle{agsm}
\spacingset{1}
\bibliography{references}


\counterwithin{lemma}{subsection}
\counterwithin{proposition}{subsection}
\counterwithin{theorem}{subsection}
\counterwithin{corollary}{subsection}
\counterwithin{assumption}{subsection}
\counterwithin{definition}{subsection}
\counterwithin{example}{subsection}
\counterwithin{remark}{subsection}

\def\spacingset#1{\renewcommand{\baselinestretch}%
{#1}\small\normalsize} \spacingset{1}

\appendix
\spacingset{1.9} 
\section*{Online Appendix}
\addcontentsline{toc}{section}{Appendices}
\renewcommand{\thesubsection}{\Alph{subsection}}

The appendix is organized as follows. In Section \ref{app: Sharpness}, we extend Proposition \ref{Equiv_PartId} from the main text to bounds of $\mathbb{E}[Y^a | A = a', L = l]$ and $\mathbb{E}(Y^g)$. In Section \ref{app: Reversal}, we provide an example of a data-generating mechanism satisfying the IV conditions where the optimal regime is not identified, but the superoptimal regime is fully identified and is different across values of $A$ as illustrated in Figure \ref{fig:Reversal main} of the main text. In Section \ref{app: BP liability judgment}, we use the data generating mechanism of the liability judgment example in \citet{BalkePearl1997} to illustrate Corollary \ref{cor: l-opt identified}. In Section \ref{app: superopt algo}, we give the algorithm that outputs the induced bounds of $h_{\textbf{opt}}(1-a,l)$ and $g_{\textbf{sup}}(a,l)$ from bounds of $\mathbb{E}[Y^a | L = l]$. In Section \ref{app: Decision-making criteria}, we give conditions for some superoptimal decision criteria to be equal and extend the mixed strategy and the optimality results in \citet{Cui2021Individualized}. In Section \ref{app: Use of superopt regime}, we restate the procedure from \citet{StensrudSarvet2022} for using the superoptimal regime without revealing the decision maker's natural treatment value. In Section \ref{app: levis procedure}, we describe the estimation procedure from \citet{levis2023covariateassisted} and give the conditions for $\sqrt{n}$-consistency of our estimator of $\mathbb{E}[Y^g]$ for any regime $g$. In Section \ref{Jiang and Ding procedure}, we restate a procedure from \citet{JiangDing2018} for estimation and inference conditional on values of $L$. In Section \ref{appendix:fixed_obs}, we give explicit definitions of the fixed and observed regimes, and restate regimes that depend on the instrument $Z$ in instrumental variable settings described by \citet{StensrudSarvet2022}. In Sections \ref{app: Details of NLSYM} and \ref{sec: Influenza regime bounds example}, we perform detailed data analyzes using our methods for the National Longitudinal Study of Young Men (NLSYM) and influenza data from \citet{McDonald1992}, respectively. In Section \ref{app: Non-binary A}, we give some ideas of extensions for a non-binary treatment $A$. In Section \ref{app: Bound widths}, we give conditions for bounds of value functions of superoptimal regimes to be tighter than bounds of optimal regimes. Proofs of propositions in the main text and the appendix are given in Section \ref{app: proofs}.
\subsection{Extension of sharpness results}
\label{app: Sharpness}

Let $a  = 1-a' \in\{0,1\}$. By Lemma \ref{lemmma1} in the main text,
\begin{align*}
    \mathbb{E}(Y^a | A = a', L = l) = \frac{\mathbb{E}(Y^a|L = l) - \mathbb{E}(Y | A = a, L = l)P(A = a,L = l)}{P(A = a', L =l)}.
\end{align*}
Hence, 
\begin{align*}
    & \frac{\mathbf{L}_a(l) - \mathbb{E}(Y | A = a, L = l)P(A = a,L = l)}{P(A = a', L =l)} \leq \mathbb{E}(Y^a | A = a', L = l) \\
    &\leq \frac{\mathbf{U}_a(l) - \mathbb{E}(Y | A = a, L = l)P(A = a,L = l)}{P(A = a', L =l)}.
\end{align*}
Thus, the width of the bounds of $\mathbb{E}(Y^a | A = a', L = l)$ is equal to the width of the bounds of the $(l,a')$-CATE, because, under consistency, $\mathbb{E}(Y^a | A = a, L = l) = \mathbb{E}(Y | A = a, L = l)$. We argue by contradiction, as in the proof of Proposition \ref{Equiv_PartId} in online Appendix \ref{app: Proof of Equiv_PartId}, to show that the bounds of $\mathbb{E}(Y^a | A = a', L = l)$ are sharp if and only if the bounds of $\mathbb{E}(Y^a|L = l)$ are sharp.
\subsubsection{Relation to sharp bounds on $\mathbb{E}(Y^a | L = l)$}
\label{app: Necessary sharp bounds}
The following proposition shows that sharp bounds of $\mathbb{E}(Y^a | L = l)$ are necessary for the construction of sharp bounds of $\mathbb{E}(Y^g)$.
\begin{proposition}
    Let $\mathbf{L}^g \leq \mathbb{E}(Y^g) \leq \mathbf{U}^g$, where $g$ is a function of $A \in \{0,1\}$ and $L \in \mathcal{L}$. Then,
    \begin{align*}
        &\frac{\mathbf{L}^{g_{a,l}} - \sum_{l' \neq l, a'} \mathbb{E}(Y | A = a', L = l')P(A = a',L = l)}{P(L = l)} \leq\\
        & \mathbb{E}(Y^a | L = l) \leq \frac{\mathbf{U}^{g_{a,l}} - \sum_{l' \neq l, a'} \mathbb{E}(Y | A = a', L = l')P(A = a',L = l)}{P(L = l)}
    \end{align*}
    for any $a,l$ provided that $P(L = l) > 0$, where
    \begin{equation*}
        g_{a,l}(A,L) := \begin{cases} 
    a & \text{if } L = l. \\
    A & \text{if } L \neq l.
\end{cases}.
    \end{equation*}
    
    The width $w(a,l)$ of the bounds on $\mathbb{E}(Y^a | L = l)$ is
    \begin{equation*}
        w(a,l) = \frac{1}{P(L = l)}w^{g_{a,l}},
    \end{equation*}
    where $w^g := \mathbf{U}^g - \mathbf{L}^g$ is the width of the bounds of $\mathbb{E}(Y^g)$.
    \label{Arbitr_reg}
\end{proposition}

\subsection{Details on the example given in Figure \ref{fig:Reversal main}}
\label{app: Reversal}
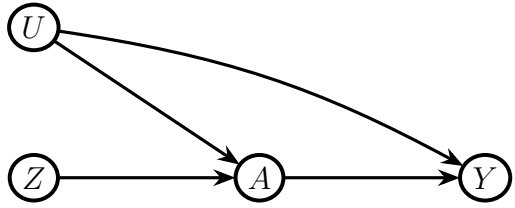
\begin{figure}
            \centering
            \begin{tikzpicture}
                \tikzset{line width=1.5pt, outer sep=0pt,
                ell/.style={draw,fill=white, inner sep=2pt,
                line width=1.5pt},
                swig vsplit={gap=5pt,
                inner line width right=0.5pt},
                swig hsplit={gap=5pt}
                };
                      \node[name=A,ell,shape=ellipse] at (6,0){$A$};
                    \node[name=Y, ell, shape=ellipse] at (9,0){$Y$};
                    \node[name=U, ell, shape=ellipse] at (3,2){$U$};
                    \node[name=Z, ell, shape=ellipse] at (3,0){$Z$};
                \begin{scope}[>={Stealth[black]},
                  every node/.style={fill=white,circle},
                  every edge/.style={draw=black,very thick}]
                    \path[->] (U) edge[bend left=10] (Y);
                    \path[->] (U) edge (A);
                    \path[->] (Z) edge (A);
                    \path[->] (A) edge (Y);
                \end{scope}
            \end{tikzpicture}
            \caption{DAG describing an instrumental variable setting.}
                            \label{fig:IV}
        \end{figure}

\begin{figure}
    \centering
    \includegraphics[scale = 0.4]{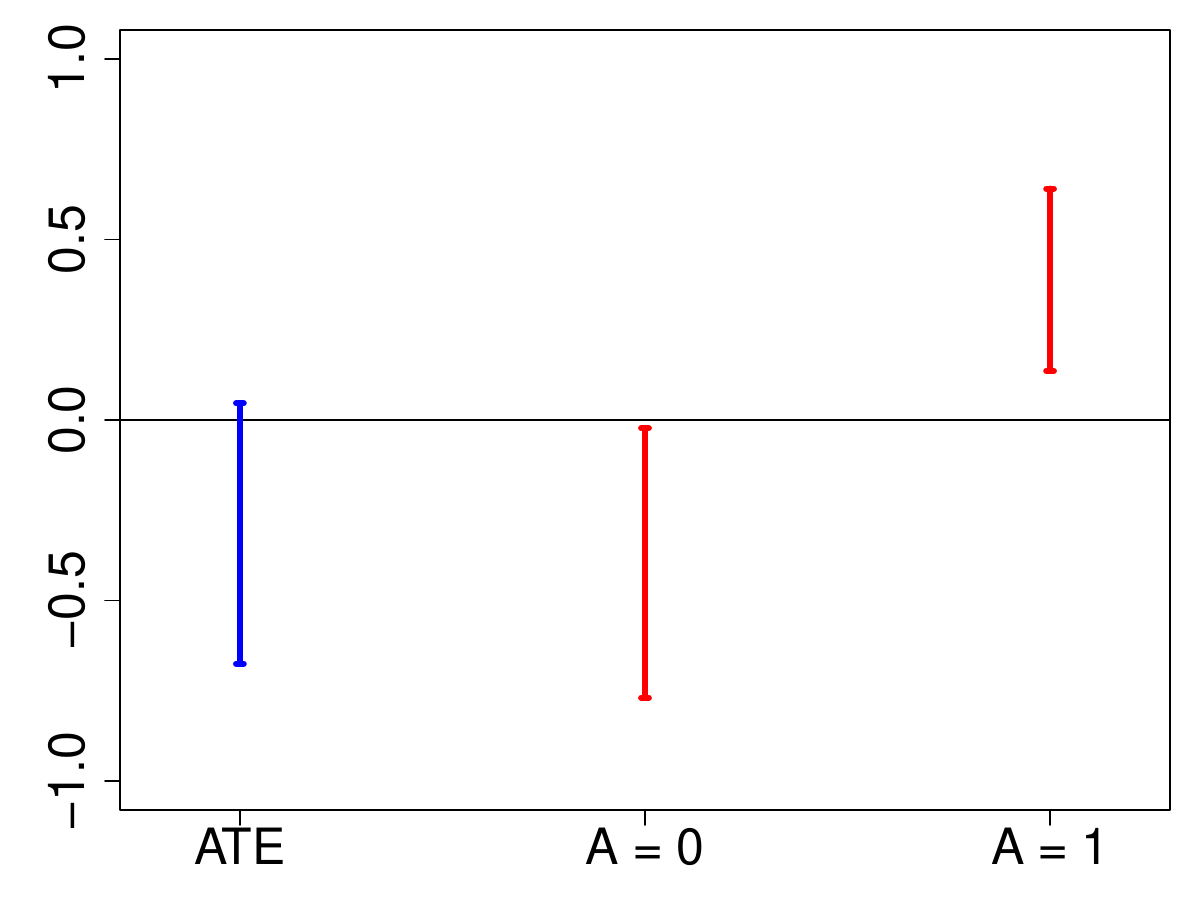}
    \caption{Bounds on the CATEs for the example of online Appendix \ref{app: Reversal}, as also presented in Figure \ref{fig:Reversal main} of the main text.}
    \label{fig:Reversal}
\end{figure}

Consider an instrumental variable setting as illustrated by the directed acyclic graph in Figure \ref{fig:IV}.
Here we will present an example, satisfying a conventional binary instrumental variable model, where the $(l,a')$-optimal and $(l,1-a')$-optimal regimes have different signs, that is,  $\sign(\mathbb{E}(Y^{a = 1} - Y^{a = 0} )) = \sign(\mathbb{E}(Y^{a = 1} - Y^{a = 0}| A = a')) = - \sign(\mathbb{E}(Y^{a = 1} - Y^{a = 0} | A = 1 - a'))$. Moreover, the example verifies that the bounds $\mathbf{L} \leq \sign(\mathbb{E}(Y^{a = 1} - Y^{a = 0} )) \leq \mathbf{U}$, $\mathbf{L}(a') \leq \sign(\mathbb{E}(Y^{a = 1} - Y^{a = 0}| A = a')) \leq \mathbf{U}(a')$ and $\mathbf{L}(1 - a') \leq \sign(\mathbb{E}(Y^{a = 1} - Y^{a = 0}| A = 1 - a')) \leq \mathbf{U}(1 - a')$ verify $\mathbf{L} \leq 0 \leq \mathbf{U}$ and $\mathbf{U}(a') < 0 < \mathbf{L}(1 - a')$, i.e., although the bounds of the ATE cover zero, we can identify the superoptimal regimes $g_{\textbf{sup}}(a') = 0 = 1 - g_{\textbf{sup}}(1-a')$ (see Figure \ref{fig:Reversal}).

\citet{BalkePearl1997} introduced variables $r_a$ and $r_y$ such that:
\begin{enumerate}
    \item $r_a = 0 \Rightarrow A^z = 0$
    \item $r_a = 1 \Rightarrow \begin{cases}
        A^{z = 0} = 0\\
        A^{z = 1} = 1
    \end{cases}$
    \item $r_a = 2 \Rightarrow \begin{cases}
        A^{z = 1} = 0\\
        A^{z = 0} = 1
    \end{cases}$
    \item $r_a = 3 \Rightarrow A^z = 1$
\end{enumerate}
and
\begin{enumerate}
    \item $r_y = 0 \Rightarrow Y^a = 0$
    \item $r_y = 1 \Rightarrow \begin{cases}
        Y^{a = 0} = 0\\
        Y^{a = 1} = 1
    \end{cases}$
    \item $r_y = 2 \Rightarrow \begin{cases}
        Y^{a = 1} = 0\\
        Y^{a = 0} = 1
    \end{cases}$
    \item $r_y = 3 \Rightarrow Y^a = 1$
\end{enumerate}
Then, the instrumental variable graph in Figure \ref{fig:IV} is equivalent to the graph in Figure \ref{fig:BP_IV} \citep{BalkePearl1997}. The joint distribution of $r_a$ and $r_d$ is given in our example in Figure \ref{fig:p_ra.ry}.

\begin{figure}
            \centering
            \begin{tikzpicture}
                \tikzset{line width=1.5pt, outer sep=0pt,
                ell/.style={draw,fill=white, inner sep=2pt,
                line width=1.5pt},
                swig vsplit={gap=5pt,
                inner line width right=0.5pt},
                swig hsplit={gap=5pt}
                };
                    \node[name=A, ell, shape=ellipse] at (6,0){$A$};
                    \node[name=Y, ell, shape=ellipse] at (9,0){$Y$};
                    \node[name=r_d, ell, shape=ellipse] at (5,2){$r_a$};
                    \node[name=Z, ell, shape=ellipse] at (3,0){$Z$};
                    \node[name = r_y, ell, shape=ellipse] at (7,1){$r_y$};
                \begin{scope}[>={Stealth[black]},
                  every node/.style={fill=white,circle},
                  every edge/.style={draw=black,very thick}]
                    \path[->] (r_d) edge (r_y);
                    \path[->] (r_d) edge (A);
                    \path[->] (r_y) edge (A);
                    \path[->] (r_y) edge (Y);
                    \path[->] (Z) edge (A);
                    \path[->] (A) edge (Y);
                \end{scope}
            \end{tikzpicture}
            \caption{Instrumental variable graph, which is a modified version of Figure 3 in \citet{BalkePearl1997}, where we omit $r_z$.}
                            \label{fig:BP_IV}
\end{figure}
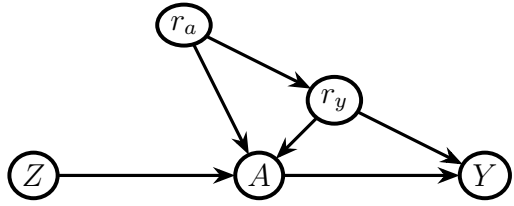

\begin{figure}
    \centering
    \begin{tabular}{ |p{1.25cm}||p{1.5cm}|p{1.5cm}|p{1.5cm}|p{1.5cm}|p{1.5cm}| }
 \hline
  & $r_y = 0$ & $r_y = 1$ & $r_y = 2$ & $r_y = 3$\\
 \hline
 $r_a = 0$ & 0.0612 & 0.004 & 0.439 & 0.081\\
 $r_a = 1$ & 0.011 & 0.010 & 0.1291 & 0.162\\
 $r_a = 2$ & 0.020 & 0.051 & 0.011 & 0.015\\
 $r_a = 3$ & 0.002 & 0.001 & 0.001 & 0.000\\
 \hline
\end{tabular}
    \caption{Joint distribution of $r_a$ and $r_y$. We consider the instrument $Z$ to be a binary random variable with success probability $P(Z = 1) = 0.008$.}
    \label{fig:p_ra.ry}
\end{figure}

\begin{figure}
    \centering
    \begin{tabular}{ |p{4cm}||p{1.5cm}| }
 \hline
  Quantity & Value \\
 \hline
 $P(A = 1)$ & $0.104$\\
 $P(Z = 1)$ &  $0.008$ \\
 $P(A = 1 |Z = 0) $ & $ 0.102$ \\
 $P(A = 1 | Z = 1) $ & $ 0.316$ \\
 $\mathbb{E}(Y^{a = 1} - Y^{a = 0})  $ & $ -0.514$\\
 $\mathbb{E}(Y^{a = 1} - Y^{a = 0} | A = 0) $ & $ -0.617$\\
 $\mathbb{E}(Y^{a = 1} - Y^{a = 0} | A = 1) $ & $ 0.369$\\
 \hline
\end{tabular}
    \caption{Some quantities of interest in the joint distribution of $Y$, $A$ and $Z$. As we are only interested in giving an intuition about the distribution in Figure \ref{fig:p_ra.ry}, we round these numbers up to three decimal places.}
    \label{fig:key_nums}
\end{figure}

Here we give an (intuitive) explanation why the distribution described in Figure \ref{fig:p_ra.ry} satisfies the statement of Corollary \ref{noCATE_Id} in the main text. The effect of $Z$ on $A$ needs to be large enough to obtain bounds of $\mathbb{E}(Y^{a = 1} - Y^{a = 0} | A = a)$ of small width for $a \in \{0,1\}$. However, the width of the bounds cannot be too small as otherwise the sign of the ATE will be identified. Hence, we need a moderate effect of $Z$ on $A$ (see Figure \ref{fig:key_nums}). Furthermore, the confounding factor $U$ needs to be strong as the effects $\mathbb{E}(Y^{a = 1} - Y^{a = 0} | A = a)$ change sign depending on the value of $A$.

\subsection{\citet{BalkePearl1997} liability judgment example}
\label{app: BP liability judgment}

Consider the fictional liability judgment example of a marketer that randomly mails out product samples of a medication that can cause peptic ulcers to households in a city\citep{BalkePearl1997}. Define $Z$ to be be an indicator function of the receipt of the sample, $A$ to be an indicator of consumption of the sample, and $Y$ to be an indicator of peptic ulcers in the following month,  and consider a specific observed distribution (Table \ref{tab:Balke-Pearl liability example}). First, set $P(Z = 1)$ to $0.1$.
\begin{table}
    \centering
    \begin{tabular}{|c|c|} \hline 
         $P(Y = 0, A = 0 \mid Z = 0) = 0.32$& $P(Y = 0, A = 0 \mid Z = 1) = 0.02$\\ \hline 
         $P(Y = 0, A = 1 \mid Z = 0) = 0.32$& $P(Y = 0, A = 1 \mid Z = 1) = 0.17$\\ \hline 
         $P(Y = 1, A = 0 \mid Z = 0) = 0.04$& $P(Y = 1, A = 0 \mid Z = 1) = 0.67$\\ \hline 
         $P(Y = 1, A = 1 \mid Z = 0) = 0.32$& $P(Y = 1, A = 1 \mid Z = 1) = 0.14$ \\ \hline
    \end{tabular}
    \caption{Data from the liability judgment example in \citet{BalkePearl1997}.}
    \label{tab:Balke-Pearl liability example}
\end{table}
 The sharp bounds of the $l$-CATE and the $(l,0)$ and $(l,1)$-CATEs do not cover $0$ when $P(Z = 1) = 0.1$. When $P(Z = 1)$ is changed to $0.25$, the bounds of the $l$-CATE and $(l,0)$-CATE do not cover zero, but the bounds of the $(l,1)$-CATE do. See \cite{BalkePearl1997} and online Appendix \ref{app: BP bounds formulas} for details on the sharp bounds.

\subsection{Superoptimal regime algorithm}
\label{app: superopt algo}
See Algorithm \ref{Algo: (L,A)-CATE bounds and superopt}.
\begin{algorithm}
    \KwData{ (Sharp) Bounds $\mathbf{L}_a(l) \leq \mathbb{E}(Y^a | L = l) \leq \mathbf{U}_a(l)$.}
    \KwResult{(Sharp) Bounds $\mathbf{L}(1-a,l) \leq h_{\mathbf{opt}}(1-a,l) \leq \mathbf{U}(1-a,l)$ and $g_{\textbf{sup}}(1-a,l)$}
    Compute $\mathbb{E}(Y | L=l)$\;
    Compute $P(A = 1 - a | L = l)$\;
    Plug values for $\mathbb{E}(Y | L=l)$, $P(A = 1 - a | L = l)$, and $(\mathbf{L}_a(l),\mathbf{U}_a(l))$ as described before Proposition \ref{Equiv_PartId} to obtain $(\mathbf{L}(l, 1-a ), \mathbf{U}(l, 1-a))$\;
    \eIf{$\mathbf{L}(1-a,l) > 0$}{
    $g_{\textbf{sup}}(1-a,l) \gets 1$\;}{
    \eIf{$\mathbf{U}(1-a,l) < 0$}{
    $g_{\textbf{sup}}(1-a,l) \gets 0\;$
    }{
    $g_{\textbf{sup}}(1-a,l) \gets \text{'unknown'}\;$
    }}
    Return $(\mathbf{L}(1-a,l), \mathbf{U}(1-a,l), g_{\textbf{sup}}(1-a,l)).$\ 
    \caption{$(l,1-a)$-CATE bounds and sign of $g_{\textbf{sup}}$.}
    \label{Algo: (L,A)-CATE bounds and superopt}
\end{algorithm}

\subsection{Decision criteria}
\label{app: Decision-making criteria}

When studying $(L,A)$-optimal regimes, the functions $g, \mathbf{L}, \mathbf{U}, \mathbf{L}_a$ and $\mathbf{U}_a$ are re-defined to be functions of the natural treatment value, see Proposition \ref{Equiv_PartId} in the main text and online Appendix \ref{app: Sharpness}. It is easy to verify the $(L,A)$-optimal decision criteria minimize a weighted risk, analogously to the $L$-optimal decision criteria in \cite{Cui2021Individualized} (see Table \ref{fig:Decision-making criteria} in the main text), which we show in online Appendix \ref{App: Fct optimization property for (L,A)-optimal decision criteria}.
\begin{proposition}
 The healthcare decision making criterion with $a = 0$ baseline is $I(\mathbf{L}(l, a') > 0)$ when $A = a'$. This criterion coincides with the optimist if $A=1$ and the pessimist if $A=0$. 
\label{Criteria_Equiv}
\end{proposition}
\cite{Cui2021Individualized} presented a mixed strategy where the probability of selecting treatment $a = 1$ is randomized and guaranteed to outperform the opportunist criterion in expectation. Here we confirm that a similar guarantee holds in the $(L,A)$-optimal setting. 
\begin{proposition}
Consider the mixed strategy for decision-making \citep{Cui2021Individualized}, where treatment $a = 1$ is chosen with probability $p(l,a')$ defined by the minimax regret criterion
\begin{equation*}
    \min_{p(l,a')}\max\{[1-p(l,a')]\max\{\mathbf{U}(l,a'),0)\}, p(l)\max\{-\mathbf{L}(l,a'),0\}\}.
\end{equation*}
This strategy is better in expectation than the opportunist strategy when $A = a', L = l$.
\label{Mixed_and_Opp}
\end{proposition}

The mixed strategy criterion using $A$ may not be better than the analogous one that only depends on $L$ in \cite{Cui2021Individualized}. Indeed, as it is based on worst case regret, the mixed strategy stratified across $A = 0,1$ may overestimate the true regret for some value of $A$, unlike the mixed strategy from \cite{Cui2021Individualized} that is not a function of $A$. 

\subsubsection{A property of $(L,A)$-optimal decision criteria}
\label{App: Fct optimization property for (L,A)-optimal decision criteria}
We follow the derivations in \cite{Cui2021Individualized}, but we add $A$ as input to the risk and weight functions. Let $\mathcal{D}^{\text{sup}}(a')$ be the set of regimes $g$ that maximize the following lower bound of $\mathbb{E}(Y^g | A = a')$ for $a' = 0,1$,
\begin{equation}
\begin{split}
    \mathbb{E}_{L}\{&[1 - w(L,a')][\mathbf{L}(a',L)I(g(a',L) = 1) + \mathbf{L}_0(a',L)] \\
    &+ w(L,a')[-\mathbf{U}(a',L)I(g(a',L) = 0) + \mathbf{L}_1(a',L)]\},
\end{split}
    \label{eq: objective function}
\end{equation}
where $\mathbf{L}_{a}(A,L) \leq \mathbb{E}(Y^{a} | A, L) \leq \mathbf{U}_{a}(A,L)$ are the induced bounds. The weight function $w(l,a')$ can depend on $g$, and $0 \leq w(l,a') \leq 1$ for any $(l,a') \in \mathcal{L} \times \{0,1\}$.
 We define $P := \mathbf{L}(a',L)I(g(a',L) = 1) + \mathbf{L}_0(a',L)$ and $Q := -\mathbf{U}(a',L)I(g(a',L) = 0) + \mathbf{L}_1(a',L)$ \citep{Cui2021Individualized}. Then, the optimist decision criterion is the regime that minimizes the objective function in Equation \eqref{eq: objective function} for weight function $w(l,a') = I(P > Q)$. Similarly, the pessimist, opportunist and healthcare decision criteria (with baseline treatment $a = 0$) correspond to weight functions $I(P < Q), 1/2$ and $0$ respectively.

If $0 \notin (\mathbf{L}(a',l), \mathbf{U}(a',l))$, regardless what weight function $w$ is chosen, the regime $g$ maximizing objective function \eqref{eq: objective function} matches the $(l,a')$-optimal regime, i.e. $g(a',l) = g_{\textbf{sup}}(a',l)$.

\subsection{Suggestion for use of $(L,A)$-optimal regimes}
\label{app: Use of superopt regime}
\citet{StensrudSarvet2022} suggested the following procedure for using the natural treatment value in practice.

First, output $g_{\textbf{sup}}(a',l)$ for both values of $a' \in \{0,1\}$. Then, the decision maker can use the result of the $(L,A)$-optimal regime without revealing their own. This procedure informs the decision maker if their natural treatment value is aligned with the output of the $(L,A)$-optimal regime, while keeping the natural treatment value private. If the $(L,A)$-optimal regime's output does not align with their natural treatment value, the decision maker may reconsider their decision. 

 The guarantees of the $(L,A)$-optimal regime no longer hold if the decision maker does not always comply. However, the added information provided can still improve outcomes when followed, as seen for the conventionality criterion in Section \ref{sec: ICU example}.

\subsection{\citet{levis2023covariateassisted} estimation procedure}
\label{app: levis procedure}
\subsubsection{Balke-Pearl bounds}
\label{app: BP bounds formulas}
For a given $l$, let $p_{ya.z}(l) := P(Y = y, A = a |Z = z, L = l)$, which we write $p_{ya.z}$ in the following equations for brevity. Following \citet{BalkePearl1997}, we can define
\begin{align}
    \mathbf{L}_0(l) &:= \max \{p_{10.1}, p_{10.0}, p_{10.0} + p_{11.0} - p_{00.1} - p_{11.1}, p_{01.0} + p_{10.0} - p_{00.1} - p_{01.1}\}, \label{eq: BP lower 0}\\
    \mathbf{U}_0(l) &:= \min \{1 - p_{00.1}, 1 - p_{00.0}, p_{01.0} + p_{10.0} + p_{10.1} + p_{11.1},p_{10.0} + p_{11.0} + p_{01.1} + p_{10.1} \},\label{eq: BP upper 0}\\
    \mathbf{L}_1(l) &:= \max \{p_{11.0}, p_{11.1}, p_{00.1} + p_{11.1} - p_{00.0} - p_{01.0}, p_{10.1} + p_{11.1} - p_{01.0} - p_{10.0}\},\label{eq: BP lower 1}\\
    \mathbf{U}_1(l) &:= \min \{1 - p_{01.1}, 1 - p_{01.0}, p_{00.0} + p_{11.0} + p_{10.1} + p_{11.1}, p_{10.0} + p_{11.0} + p_{00.1} + p_{11.1} \}. \label{eq: BP upper 1}
\end{align}

We first adapt the Margin condition of \citet{levis2023covariateassisted}.

\begin{assumption}[Margin condition (\cite{levis2023covariateassisted})]
    Let the Balke-Pearl bounds $\mathbf{L}(l)$ and $\mathbf{U}(l)$ be given by
    \begin{align}
        \mathbf{L}(l) &= \max_j \theta^{\mathbf{L}}_j(l) = \theta^{\mathbf{L}}_{d^\mathbf{L}(l)}(l)\text{ and }
        \mathbf{U}(l) = \min_j \theta^{\mathbf{U}}_j(l) = \theta^{\mathbf{U}}_{d^\mathbf{U}(l)}(l)
        \label{eq: Margin bounds}
    \end{align}
    where the $\theta^{\mathbf{L}}_j(l)$, $\theta^{\mathbf{U}}_j(l)$ are the appropriate sums and subtractions of the $p_{ya.z}(l) := P(Y = y, A = a | Z = z, L = l)$ and constants (see online Appendix \ref{app: BP bounds formulas}).
    
    Then, there exists $\alpha > 0$ such that for any $t \geq 0$,
    \begin{equation*}
        P\left[\min_{j \neq d^{\mathbf{L}}(L)} \{\theta^{\mathbf{L}}_j(L) - \theta^{\mathbf{L}}_{d^{\mathbf{L}}(L)}(L)\} \leq t \right] \lesssim t^{\alpha},
        \text{ and }
        P\left[\min_{j \neq d^{\mathbf{U}}(L)} \{\theta^{\mathbf{U}}_j(L) - \theta^{\mathbf{U}}_{d^{\mathbf{U}}(L)}(L)\} \leq t \right] \lesssim t^{\alpha},
    \end{equation*}
    where $\lesssim$ represents that the inequality holds up to a multiplicative constant.
    \label{As:Margin}
\end{assumption}
This assumption controls the probability that the $\theta_j$ are close to their extrema given by indices $d^{\mathbf{L}}(l)$ and $d^{\mathbf{U}}(l)$. Similar conditions have been used in classification problems (\cite{Audibert2007}), dynamic treatment regimes (\cite{Luedtke2016}), and other IV settings (\cite{Kennedy2020sharp_instruments}). The $\theta_j(l)$ are sums and subtractions of observable joint probabilities $p_{ya.z}(l) = P(Y = y, A = a | Z = z, L = l)$. \cite{levis2023covariateassisted} argued that Assumption \ref{As:Margin} is plausible because violating Assumption \ref{As:Margin} seems to necessitate complex and unlikely relationships between the $p_{ya.z}(l)$ for several of the $\theta_j(l)$ to be the sharpest lower or upper bounds on our estimands.

Under Assumption \ref{As:Margin} and other regularity assumptions given in Theorem \ref{thm: Levis} in online Appendix \ref{app: Levis theorem}, \cite{levis2023covariateassisted} showed they can tighten the Balke-Pearl bounds using observed covariates. Specifically, for every $y,a$ and $z \in \{0,1\}$ let
\begin{equation*}
    \psi_{ya.z}(L) = \frac{I(Z = z)}{\lambda_z(L)}\{I(Y = y, A = a) - p_{ya.z}(L)\}
\end{equation*}
be the influence function of $p_{ya.z}(l) := P(Y = y, A = a | Z = z, L = l)$, where $\lambda_z(L) = P(Z = z | L)$. \cite{levis2023covariateassisted} assumed that there exists $\epsilon > 0 $ such that $\lambda_1(L) \in [\epsilon, 1-\epsilon]$ almost surely, and we will use this assumption henceforth. 

Then, let $\mathbf{L}_j(L)$ be obtained by taking the formula for $\theta_j^{\mathbf{L}}(L)$, replacing $p_{ya.z}$ with $\psi_{ya.z}$ and omitting additive constants, such that $\mathbf{L}_j(L)$ is the influence function of $\theta_j^{\mathbf{L}}(L)$. Analogously, let $\mathbf{U}_j(L)$ be obtained by taking the formula for $\theta_j^{\mathbf{U}}(L)$, replacing the $p_{ya.z}$ by the $\psi_{ya.z}$ and omitting additive constants. Then, similarly, $\mathbf{U}_j(L)$ is the influence function of $\theta_j^{\mathbf{U}}(L)$. The bounds in Equation \eqref{eq: Margin bounds} verify
\begin{align*}
    \mathbf{L}(l) &= \sum_j I(d^{\mathbf{L}}(l) = j)\theta^{\mathbf{L}}_j(l), \text{ and }
    \mathbf{U}(l) = \sum_j I(d^{\mathbf{U}}(l) = j)\theta^{\mathbf{U}}_j(l). 
\end{align*}
The bounds $\mathbf{L}(l)$ and $\mathbf{U}(l)$ have the uncentered influence functions
\begin{align*}
    \varphi^{\mathbf{L}}_{d^{\mathbf{L}}}(L) &= \sum_j I\{ d_{\mathbf{L}}(L) = j \}\{\mathbf{L}_j(L) + \theta^{\mathbf{L}}_j(L) \}, \quad 
    \varphi^{\mathbf{U}}_{d^{\mathbf{U}}}(L) = \sum_j I\{ d_{\mathbf{U}}(L) = j \}\{\mathbf{U}_j(L) + \theta^{\mathbf{U}}_j(L) \}.
\end{align*}
The estimation procedure in \cite{levis2023covariateassisted} is generically provided in Algorithm \ref{algo:levis bounds}. 
\begin{algorithm}
    \KwData{Observations $(A_i,Z_i,L_i,Y_i)_i$}
    \KwResult{Estimates of bounds $\mathbf{L}$ and $\mathbf{U}$}
    Estimate $p_{ya.z}$ and deduce an estimator $\hat{\theta}_j$ for $\theta_j$\;
    Estimate $d^{\mathbf{L}}$ and $d^{\mathbf{U}}$ by $\hat{d^{\mathbf{L}}}(l) = \arg \max_j \hat{\theta}^{\mathbf{L}}_j(l)$ and $\hat{d^{\mathbf{U}}}(l) = \arg \min_j \hat{\theta}^{\mathbf{U}}_j(l)$\;
    Estimate the $\lambda_z$ and deduce an estimate of the $\psi_{ya.z}$ using the estimators for the $p_{ya.z}$. Use these estimates to estimate $\mathbf{L}_j$ and $\mathbf{U}_j$\;
    Estimate $\mathbf{L}$ and $\mathbf{U}$ as 
    \begin{align}
        \hat{\mathbf{L}} &:= \sum_j \mathbb{P}_n [I(\hat{d}^{\mathbf{L}}(L) = j)\{\hat{\mathbf{L}}_j(L) + \hat{\theta}^{\mathbf{L}}_j(L)\}] = \mathbb{P}_n[\hat{\varphi}^{\mathbf{L}}_{\hat{d}^{\mathbf{L}}}(L)], \label{eq: Levis lower estimate}\text{ and} \\
        \hat{\mathbf{U}} &:= \sum_j \mathbb{P}_n [I(\hat{d}^{\mathbf{U}}(L) = j)\{\hat{\mathbf{U}}_j(L) + \hat{\theta}^{\mathbf{U}}_j(L)\}] = \mathbb{P}_n[\hat{\varphi}^{\mathbf{U}}_{\hat{d}^{\mathbf{U}}}(L)]\; \label{eq: Levis upper estimate}
    \end{align}
    \caption{Algorithm for computing bounds, adapted from \cite{levis2023covariateassisted}.}
    \label{algo:levis bounds}
\end{algorithm}

Using Assumption \ref{As:Margin}, Theorem \ref{thm: Levis} in online Appendix \ref{app: Levis theorem} gives guarantees for the estimation strategy in Algorithm \ref{algo:levis bounds}. Furthermore, this theorem gives us conditions for asymptotic linearity of $\mathbf{L}$ and $\mathbf{U}$, e.g.\ suggesting the use of a bootstrap procedure for inference \citep{Davison_Hinkley_1997}. Asymptotic linearity is dependent on Assumption \ref{As:Margin} to remove discontinuity in the asymptotic distribution. Without Assumption \ref{As:Margin}, using a bootstrap procedure will be invalid because of discontinuities in the asymptotic distribution due to the $\min, \max$ functions \citep{Andrews2000, Chernozhukov2007, Romano2008, Andrews2009, tamer_partial_2010}. Alternatively, we can use standard errors for influence functions to derive confidence intervals, as suggested by a reviewer. We use bootstrap samples to compute an estimate for the variance of our estimators and then construct Wald-type confidence intervals \citep{levis2023covariateassisted}. In the next section, we adapt these results to estimating $(L,A)$-Optimal regimes. 
\begin{remark}
    Although \citet{levis2023covariateassisted} focused on $l$-CATEs and ATEs, the Balke-Pearl bounds in Assumption \ref{As:Margin} can also refer to Balke-Pearl bounds of the value functions $\mathbb{E}(Y^a | L = l)$. Provided Assumption \ref{As:Margin} holds for the appropriate bounds, Theorem \ref{thm: Levis} is true for $\mathbf{L} \leq \mathbb{E}(Y^a) \leq \mathbf{U}$ or $\mathbf{L} \leq \mathbb{E}(Y^{a = 1} - Y^{a = 0}) \leq \mathbf{U}$.
\end{remark}
\subsubsection{Estimation Theorem from \citet{levis2023covariateassisted}}
\label{app: Levis theorem}
\begin{theorem}[\cite{levis2023covariateassisted}]
Suppose that we estimate $p_{ya.z}$ and $\lambda_z$ in different samples using Algorithm \ref{algo:levis bounds}. Suppose Assumption \ref{As:Margin} holds and $P(\epsilon \leq  \lambda_z(L) \leq 1 - \epsilon) = 1$ for some $\epsilon > 0$. Suppose also $||\hat{\lambda}_1 - \lambda_1|| = o_P(1)$ and $\max_{y,a,z \in \{0,1\}} || \hat{p}_{ya.z} - p_{ya.z} || = o_P(1)$. Then,
\begin{equation*}
    \hat{\mathbf{L}} - \mathbf{L} = (\mathbb{P}_n - P)\varphi^{\mathbf{L}}_{d^{\mathbf{L}}} + O_P(||\hat{\lambda}_1 - \lambda_1 || \cdot \max_{y,a,z \in \{0,1\}} || \hat{p}_{ya.z} - p_{ya.z} || + \max_j ||\hat{\theta}^{\mathbf{L}}_{j} - \theta^{\mathbf{L}}_{j}||_{\infty}^{1 + \alpha}) + o_P(n^{-1/2})
\end{equation*}
and 
\begin{equation*}
    \hat{\mathbf{U}} - \mathbf{U} = (\mathbb{P}_n - P)\varphi^{\mathbf{U}}_{d^{\mathbf{U}}} + O_P(||\hat{\lambda}_1 - \lambda_1 || \cdot \max_{y,a,z \in \{0,1\}} || \hat{p}_{ya.z} - p_{ya.z} || + \max_j ||\hat{\theta}^{\mathbf{U}}_{j} - \theta^{\mathbf{U}}_{j}||_{\infty}^{1 + \alpha}) + o_P(n^{-1/2})
\end{equation*}
\label{thm: Levis}
\end{theorem}
This theorem shows that we get $\sqrt{n}$-convergence of our estimators of the bounds if
\begin{align*}
    ||\hat{\lambda}_1 - \lambda_1 || \cdot \max_{y,a,z \in \{0,1\}} || \hat{p}_{ya.z} - p_{ya.z} || + \max_j ||\hat{\theta}^{\mathbf{L}}_{j} - \theta^{\mathbf{L}}_{j}||_{\infty}^{1 + \alpha} = o_P(n^{-1/2}).
\end{align*}
\citet{levis2023covariateassisted} argue that usually $\max_{y,a,z \in \{0,1\}} || \hat{p}_{ya.z} - p_{ya.z} || = o_P(n^{-1/4})$ is sufficient to imply $\max_j ||\hat{\theta}^{\mathbf{L}}_{j} - \theta^{\mathbf{L}}_{j}||_{\infty}^{1 + \alpha} = o_P(n^{-1/2})$.

\subsubsection{Convergence of one-step estimator}
\label{app: One-step estimator convergence}

Let
\begin{align*}
    \Psi(a,l) &= \sum_j I(d^{\Psi}(a,l) = j)\Psi_j(g(a,l)),
\end{align*}
where
\begin{align*}
    \Psi_j(a,l) &= I(g(a,l) = a)\mathbb{E}[Y | A = a, L = l]\\
    &+ I(g(a,l) \neq a) \frac{\theta_j(l) - \mathbb{E}[Y | A = 1-a, L = l] P(A = 1-a | L = l)}{P(A = a | L = l)},\\
    d^{\Psi}(a,l) &= \arg \max_j \Psi_j(g(a,l)).
\end{align*}

Furthermore, let 
\begin{align*}
     \hat{\Psi} &:= \sum_j \mathbb{P}_n [I(\hat{d}^{\Psi}(L) = j)\{\hat{\Psi}^{\text{eff}}_j(A,L) + \hat{\Psi}_j(A,L)\}] = \mathbb{P}_n(\hat{\Psi}_{\hat{d}^{\Psi}} + \hat{\Psi}_{\hat{d}^{\Psi}}^{\text{eff}}),
\end{align*}
where $\Psi_j^{\text{eff}}(A,L)$ is the influence function of $\Psi_j(A,L)$ given in Proposition \ref{thm:proof eff infl}.

\begin{proposition}
Suppose that we estimate $p_{ya.z}$ and $\lambda_z$ in different samples using Algorithm \ref{algo:levis bounds}. Suppose Assumption \ref{As:Margin} holds, and suppose there exists $\epsilon > 0$ such that:
\begin{itemize}
    \item $P(\epsilon \leq  \lambda_z(L) \leq 1 - \epsilon) = 1$,
    \item $P(\epsilon \leq P(A = a | L ) \leq 1 - \epsilon) = 1$,
    \item $P(|Y| \leq 1/\epsilon ) = 1$,
    \item $P(\epsilon \leq \hat{P}(A = a | L) \leq 1 - \epsilon) = 1$,
    \item $P(\hat{\mathbb{E}}[Y | A = a, L] \leq 1/\epsilon ) = 1$.
\end{itemize}
 Furthermore, suppose that
 \begin{itemize}
     \item  $||\hat{\lambda}_1 - \lambda_1|| \cdot \max_{y,a,z \in \{0,1\}} || \hat{p}_{ya.z} - p_{ya.z} || = o_P(n^{-1/2})$,
     \item $\max_j ||\hat{\theta}_{j} - \theta_{j}||_{\infty}^{1 + \alpha} = o_P(n^{-1/2})$,
     \item $\hat{\mathbb{E}}[Y | A = a, L = l]$ is a $P$-Donsker estimator of $\mathbb{E}[Y | A = a, L = l]$,
     \item $\hat{P}(A = a| L =l)$ is a $P$-Donsker estimator of $P(A = a | L = l)$.
 \end{itemize}
Then,
\begin{align*}
    \hat{\Psi} - \Psi &= (\mathbb{P}_n - P)\Psi_{d^{\Psi}}^{\text{eff}} + o_P(n^{-1/2})
\end{align*}
    \label{prop: One-step convergence}
\end{proposition}
In particular, the one-step estimator is asymptotically linear and normal, suggesting the use of a bootstrap procedure for inference \citep[Chapter 2]{Davison_Hinkley_1997}.

Similarly, we can directly apply Theorem \ref{thm: Levis} to obtain the following convergence result of the one-step estimator described in Corollary \ref{cor: sign id}.
\begin{proposition}
    Suppose that we estimate $p_{ya.z}$ and $\lambda_z$ in different samples using Algorithm \ref{algo:levis bounds}, and that we estimate $\mathbb{E}(Y)$ nonparametrically. Suppose Assumption \ref{As:Margin} holds and $P(\epsilon \leq  \lambda_z(L) \leq 1 - \epsilon) = 1$ for some $\epsilon > 0$. Suppose also $||\hat{\lambda}_1 - \lambda_1|| = o_P(1)$ and $\max_{y,a,z \in \{0,1\}} || \hat{p}_{ya.z} - p_{ya.z} || = o_P(1)$. Then,
\begin{align*}
     &\hat{\mathbb{E}}(Y) + \mathbb{P}_n\left(Y - \hat{\mathbb{E}}(Y | L) - \hat{\psi}_1(a,L) - \hat{\psi}_1^{\text{eff}}(a,L) \right) - \mathbb{E}(Y ) - \psi_1(a)\\
     &= (\mathbb{P}_n - P)( Y  - \psi_1^{\text{eff}}(a)) + O_P(||\hat{\lambda}_1 - \lambda_1 || \cdot \max_{y,a,z \in \{0,1\}} || \hat{p}_{ya.z} - p_{ya.z} || + \max_j ||\hat{\theta}_{j} - \theta_{j}||_{\infty}^{1 + \alpha})\\
     &+ o_P(n^{-1/2}),
\end{align*}
where the $\theta_{j}$ correspond to the $\theta^{\mathbf{L}}_{j}$ if $\psi_1$ is a lower bounds and $\theta^{\mathbf{U}}_{j}$ if $\psi_1$ is an upper bound.
\label{prop: cor 5 convergence result}
\end{proposition}

\subsection{Inference on conditional value functions based on results from \cite{JiangDing2018}}
\label{Jiang and Ding procedure}
The procedure in \cite{levis2023covariateassisted} can be used to estimate bounds on the marginal value function. However, estimating value functions conditional on covariates $L$ and $A$, such as $\mathbb{E}(Y^a | L = l)$ and $\mathbb{E}(Y^a | A = a', L = l)$,  would also be of interest in some settings. Here we consider one approach for estimation of conditional value functions, as suggested by \cite{JiangDing2018}. This approach is based on confidence intervals first constructed in \cite{ImbensManski20004} but exploits the $\min/\max$ structure of our bounds for asymptotic coverage results.

\begin{algorithm}
    \KwData{Observations $(A_i,Z_i,L_i,Y_i)_i$ and individual characteristics $l$}
    \KwResult{$(1-\alpha)$-CI of the parameter of interest}
    Estimate the $p_{ya.z}(l)$ and deduce estimates of the $\theta_j^{\mathbf{L}}(l)$ and $\theta_j^{\mathbf{U}}(l)$\; 
    Estimate the standard errors of the $\theta_j^{\mathbf{L}}(l)$ and $\theta_j^{\mathbf{U}}(l)$ which we denote by $\hat{\sigma}_j^{\mathbf{L}}(l)$ and $\hat{\sigma}_j^{\mathbf{U}}(l)$ respectively\;
    Find the straightforward estimates for $d^{\mathbf{L}}(l)$ and $d^{\mathbf{U}}(l)$ from the estimates of $\theta_j^{\mathbf{L}}(l)$ and $\theta_j^{\mathbf{U}}(l)$ respectively\;
    Compute $C$ as the solution to:
    \begin{equation*}
        \Phi\left( C + \frac{\hat{\theta}_{\hat{d}^{\mathbf{U}}(l)}^{\mathbf{U}} - \hat{\theta}_{\hat{d}^{\mathbf{L}}(l)}^{\mathbf{L}}}{\max\{\hat{\sigma}_{\hat{d}^{\mathbf{L}}(l)}^{\mathbf{L}}, \hat{\sigma}_{\hat{d}^{\mathbf{U}}(l)}^{\mathbf{U}} \}}\right) - \Phi(-C) = 1 - \alpha 
    \end{equation*}
    where $\Phi(\cdot)$ is the cumulative distribution function of a standard normal random variable\;
    Compute the confidence interval:
    \begin{equation*}
        \text{CI}(\hat{d}^{\mathbf{L}}(l), \hat{d}^{\mathbf{U}}(l)) = [\hat{\theta}_{\hat{d}^{\mathbf{L}}(l)}^{\mathbf{L}} - C \cdot \hat{\sigma}_{\hat{d}^{\mathbf{L}}(l)}^{\mathbf{L}}, \hat{\theta}_{\hat{d}^{\mathbf{U}}(l)}^{\mathbf{U}} + C \cdot \hat{\sigma}_{\hat{d}^{\mathbf{U}}(l)}^{\mathbf{U}}]\;
    \end{equation*}
    \caption{\cite{ImbensManski20004} and \cite{JiangDing2018}}
    \label{algo: imbens-estim}
\end{algorithm}

Algorithm \ref{algo: imbens-estim} comes with the following guarantee.
\begin{theorem}[\citet{JiangDing2018}]
    If:
    \begin{enumerate}[a)]
        \item The $\theta^{\mathbf{L}}_j$ have a unique maximum value $\theta^{\mathbf{L}}_{d^{\mathbf{L}}(l)}$ and the $\theta^{\mathbf{U}}_j$ have a unique minimum value $\theta^{\mathbf{U}}_{d^{\mathbf{U}}(l)}$,
        \item  \label{Jiang_Ding Ass 2}For any $i,j$, the asymptotic distribution of $(\hat{\theta}^{\mathbf{L}}_i, \hat{\theta}^{\mathbf{U}}_j)$ is bivariate normal with means $(\theta^{\mathbf{L}}_i, \theta^{\mathbf{U}}_j)$ and estimated standard errors $(\hat{\sigma}_i^{\mathbf{L}}, \hat{\sigma}_j^{\mathbf{U}})$,
    \end{enumerate}
    Then $\text{CI}(\hat{d}^{\mathbf{L}}(l), \hat{d}^{\mathbf{U}}(l))$ has a coverage rate at least as large as $1-\alpha$ asymptotically.
    \label{Thm:JiangDing}
\end{theorem}
This theorem requires stronger assumptions than those in \cite{levis2023covariateassisted}. Indeed, instead of a Margin Condition, like Assumption \ref{As:Margin}, we assume a unique maximum/minimum value. Furthermore, it also requires asymptotically bivariate normal estimators of the $(\theta^{\mathbf{L}}_i, \theta^{\mathbf{U}}_j)$. This is reasonable for certain situations, e.g. if $L$ is empty and we are using empirical estimators or if we are using parametric models with normal residuals.

\subsubsection{$(L,A)$-optimal case} 
Theorem \ref{Thm:JiangDing} is valid for all $\theta$ that give bounds on a certain parameter of interest. However, 
\begin{align*}
    &\frac{\max_j\{\theta_j^{\mathbf{L}}(l)\} - \mathbb{E}(Y | A = a, L = l)P(A = a | L = l)}{P(A = 1-a | L = l)}\\
    &= \max_j\left \{\frac{\theta_j^{\mathbf{L}}(l) - \mathbb{E}(Y | A = a, L = l)P(A = a | L = l)}{P(A = 1-a | L = l)} \right \} \text{, and}\\
    &\frac{\min_j\{\theta_j^{\mathbf{U}}(l)\} - \mathbb{E}(Y | A = a, L = l)P(A = a | L = l)}{P(A = 1-a | L = l)} \\
    &= \min_j\left \{\frac{\theta_j^{\mathbf{U}}(l) - \mathbb{E}(Y | A = a, L = l)P(A = a | L = l)}{P(A = 1-a | L = l)} \right \}.
\end{align*}
Therefore, we can apply Theorem \ref{Thm:JiangDing} provided the terms in the $\max$ and $\min$ have estimators which are bivariate normal and consistent. In particular, if the estimator for $\mathbb{E}(Y | A = a, L = l)$ is asymptotically normal and estimators for $P(A = a' |L = l)$ are consistent for both $a' \in \{0,1\}$, then if the $\theta_j^{\mathbf{L}}(l)$ and $\theta_j^{\mathbf{U}}(l)$ satisfy the assumptions in Theorem \ref{Thm:JiangDing} we can use Algorithm \ref{algo: imbens-estim} with
\begin{align*}
    \Tilde{\theta}_j^{\mathbf{L}}(a,l) &:= \frac{\theta_j^{\mathbf{L}}(l) - \mathbb{E}(Y | A = a, L = l)P(A = a | L = l)}{P(A = 1-a | L = l)}, \text{ and} \\
    \Tilde{\theta}_j^{\mathbf{U}}(a,l) &:= \frac{\theta_j^{\mathbf{U}}(l) - \mathbb{E}(Y | A = a, L = l)P(A = a | L = l)}{P(A = 1-a | L = l)}.
\end{align*}

\subsection{Some regimes}
\label{appendix:fixed_obs}

\begin{definition}[Fixed regime]
     A fixed regime $g:\{0,1\} \times \mathcal{L} \times \{0,1\}$ to $\{0,1\}$ assigns a (non-stochastic) treatment $A^{g_+} = a$ given any number of the variables $A = a'$, $L = l$ and $Z = z$.
\end{definition}
One example of a fixed regime is the following regime:
\begin{definition}[Observed regime]
    The observed regime $g_{\textbf{obs}}$ assigns treatment $A^{g_{\textbf{obs}+}} = a$ given variables $A = a$, $L = l$ and $Z = z$ i.e., $g_{\textbf{obs}} = A$.
\end{definition}
In particular, $\mathbb{E}[Y^{g_{\textbf{obs}}}] = \mathbb{E}[Y]$. 

\begin{definition}[$(L,Z)$-Optimal regime]
The $(L,Z)$-optimal regime $g_{\textbf{opt}}$ assigns treatment $A^{g_{\textbf{opt}+}} = a$ given $L = l$ and $Z = z$ as $g_{\textbf{opt}}(l,z) \equiv \underset{a \in \{0,1\} }{\arg \max } \  \mathbb{E}(Y^{a} \mid L = l, Z = z)$.
\label{def: (L,Z)-optimal regime}
\end{definition}
The $(L,Z)$-optimal regime is always identical to the $L$-optimal regime if $Z$ is a valid instrument in an instrumental variable setting \citep{StensrudSarvet2022}.
\begin{definition}[$(L,A,Z)$-Optimal regime]
The $(L,A,Z)$-optimal regime $g_{\textbf{sup}}$ assigns treatment $A^{g_{\textbf{sup}+}} = a$ given $A = a'$, $L = l$ and $Z = z$ as $g_{\textbf{sup}}(a',l,z) \equiv \underset{a \in \{0,1\} }{\arg \max } \  \mathbb{E}(Y^{a} \mid A = a', L = l, Z = z)$.
\label{def: (L,A,Z)-optimal regime}
\end{definition}
We will use the instrument-dependent regimes of Definitions \ref{def: (L,Z)-optimal regime} and \ref{def: (L,A,Z)-optimal regime} for an analysis of the NLSYM data in online Appendix \ref{app: NLSYM analysis conditional on Z}.

\subsection{NLSYM analysis}
\label{app: Details of NLSYM}
\subsubsection{Effect of education on future earnings}
\label{sec:NLSYM data analysis}
We applied the methods in Section \ref{sec: Estimation} of the main text to study the effect of higher education on future earnings, using observational data from the National Longitudinal Study of Young Men (NLSYM) \citep{Card1993}. \cite{Cui2021Individualized} analyzed the same data to derive bounds on the $l$-CATEs to find optimal regimes. Similarly to \cite{Cui2021Individualized}, our aim is to find optimal dynamic treatment regimes for pursuing higher education that maximize future earnings, given an individual's characteristics. However, unlike \cite{Cui2021Individualized}, we will also leverage the intention to pursue higher education, that is, the natural treatment value. 

Following \cite{Wang2018} and \cite{Cui2021Individualized}, we consider education beyond high school to be the treatment $A$, and the presence of a nearby four-year college to be the IV $Z$. We dichotomize the wage at its median to obtain the outcome $Y$. The pretreatment covariate vector $L$ contains race, parents' education levels, age, and IQ scores, see Appendix \ref{app: NLSYM} for details.



\begin{figure}
    \centering
    \includegraphics[scale = 0.55]{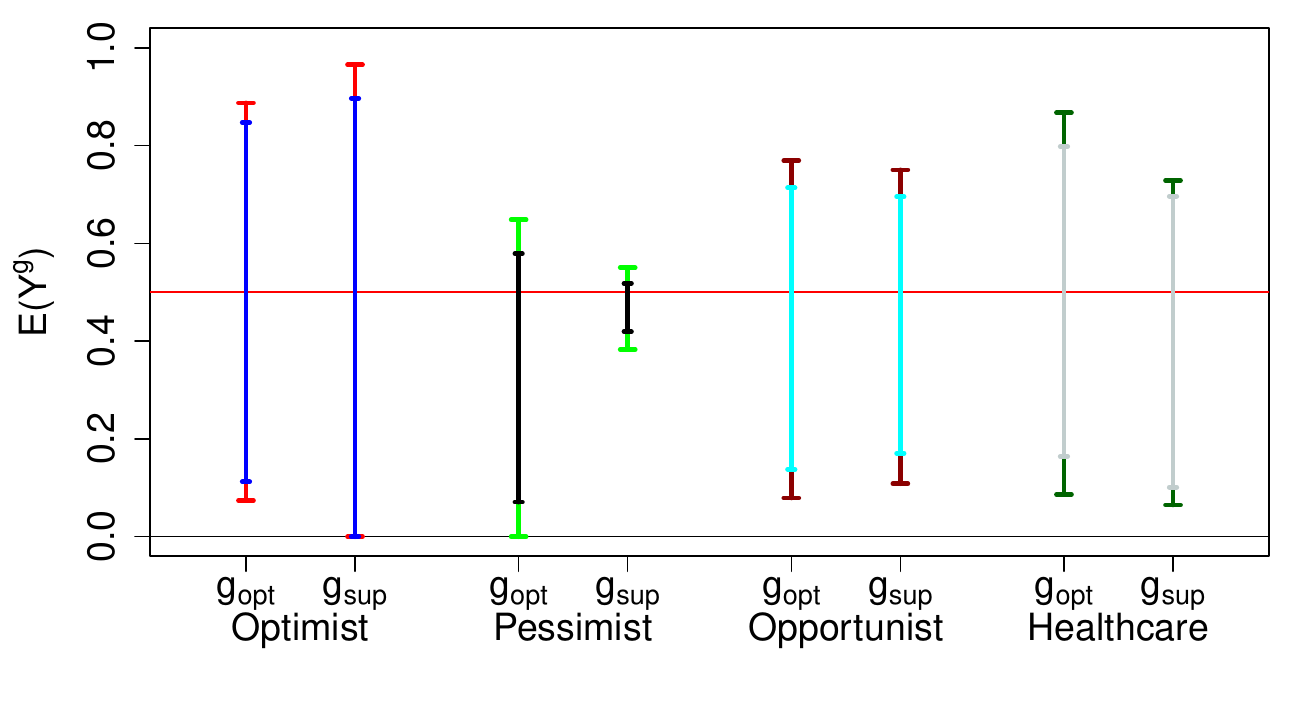}
    \caption{Bounds of value functions in the example on education and future earnings. Bounds for the estimated $L$-optimal and $(L,A)$-optimal regimes are compared. The red horizontal line represents the observed mean of $Y$, $\mathbb{E}(Y)$. 
    }
    \label{fig:Card_regime_bounds}
\end{figure}

To estimate bounds on $\mathbb{E}(Y^g)$, we did sample splitting, similar to \cite{levis2023covariateassisted}. First, the data were split $20$-$80$. On the first $20\%$ of the data, we trained parametric models for Balke-Pearl bounds $\mathbf{L}(l)$ and $\mathbf{U}(l)$ of the value functions and CATEs, which were used as input to the decision functions, as specified in Table \ref{fig:Decision-making criteria}. These results were used to find a candidate optimal regime. 
Then, on the remaining $80\%$ of the data, we performed 10-fold cross-validation to compute bounds and confidence intervals on the value function $\mathbb{E}(Y^g)$ of the decision-criteria regimes (see Figures \ref{fig:Card_regime_bounds} and \ref{fig:Influenza_regime_bounds}) under the candidate optimal regimes. The bounds on the value function under  the \textit{estimated} regimes were computed based on the strategy in Section \ref{sec: Estimation}.

The bounds are wide under all decision criteria (Figure \ref{fig:Card_regime_bounds}), and all of the bounds cover the observed mean $\mathbb{E}(Y)$ (red horizontal line in Figure \ref{fig:Card_regime_bounds}). Yet the estimated $(L,A)$-optimal regimes have narrower bounds compared to the $L$-optimal regimes for 3 out of 4 decision criteria, without imposing additional assumptions. 

\subsubsection{Additional analysis}
\begin{figure}
    \centering
    \includegraphics[scale = 0.5]{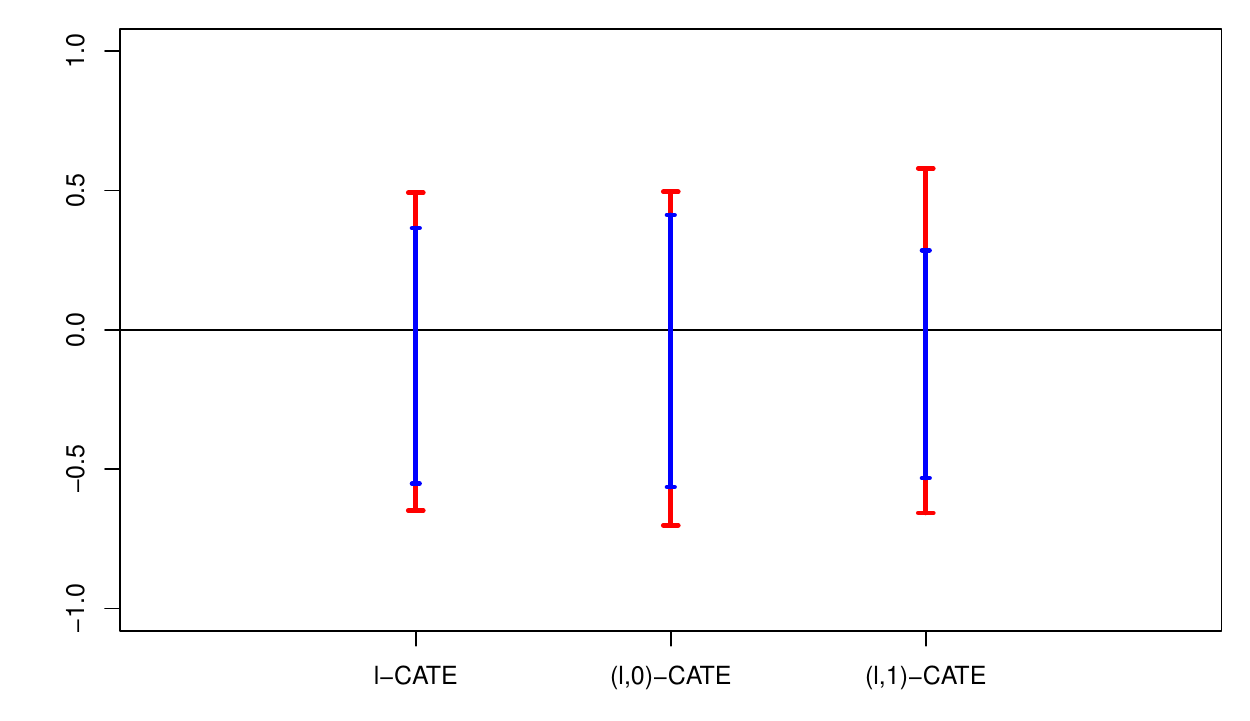}
    \caption{Bounds on the CATEs in the analysis of the NLSYM.}
    \label{fig:Card_CombinedCATE}
\end{figure}

\begin{figure}
    \centering
    \includegraphics[scale = 0.5]{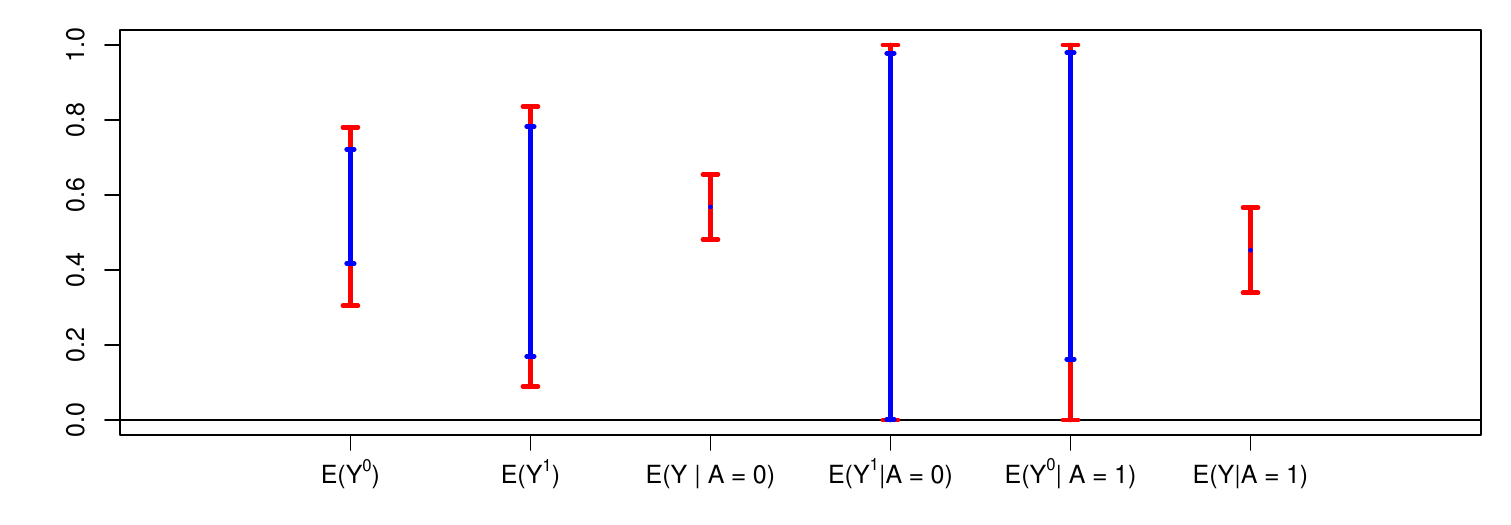}
    \caption{Estimated bounds in the analysis of the NLSYM.}
    \label{fig:Card_Estimand_bounds}
\end{figure}

We estimate the probabilities $P(Y = y, A = a| Z = z, L = l)$ using multinomial models, which are compatible with the procedure in \cite{JiangDing2018} (that we recall in online Appendix \ref{Jiang and Ding procedure}) to build 95\% confidence intervals; this procedure requires that the estimators for $\theta_j(l)$, that we maximize and minimize in order to compute the confidence intervals, are jointly normal as presented in Theorem \ref{Thm:JiangDing} \citep{JiangDing2018} given in online Appendix \ref{Jiang and Ding procedure}. If the multinomial models are correctly specified, joint normality is a consequence of a normal residuals assumption. However, these models are more restrictive than the non-parametric models of \citet{levis2023covariateassisted} that we use in Section \ref{sec: ICU example}, and we have to assume the residuals are normally distributed. If the multinomial models are credible, they permit us to conduct inference using confidence intervals on the value functions and the CATEs (Figures \ref{fig:Card_CombinedCATE} and \ref{fig:Card_Estimand_bounds}).

The bounds of the $l$-CATE cover $0$ (see Figure \ref{fig:Card_CombinedCATE}), but a decision maker could justify decisions based on explicit criteria \citep{Cui2021Individualized}. We also estimate the bounds for $\mathbb{E}(Y^a | L = l)$ (Figure \ref {fig:Card_Estimand_bounds}).

As the bounds do not identify the optimal regimes, we use the decision criteria of \cite{Cui2021Individualized} (see Figure \ref{fig:CardTableOpt}).
\begin{figure}
    \centering
    \begin{tabular}{ |p{4cm}||p{3cm}|p{3cm}|p{3cm}|p{2cm}|p{2cm}| }
 \hline
 Decision criteria & $L$-optimal & $A = 0$ & $A = 1$\\
 \hline
 Maximax(Optimist)   & $a = 1$ & $a = 1$ & $a = 0$\\
 Maximin(Pessimist) &  $a = 0$ & $a = 0$ & $a = 1$\\
 Minimax(Opportunist) & $a = 0$ & $a = 0$ & $a = 0$\\
 Healthcare decision making($a = 0$ baseline) & $a = 0$ & $a = 0$ & $a = 0$\\
 \hline
\end{tabular}
    \caption{Outcome of several decision criteria in the analysis of the NLSYM.}
    \label{fig:CardTableOpt}
\end{figure}

The bounds of the $(l,a')$-CATEs cover zero (see Figure \ref{fig:Card_CombinedCATE}), however we can still use some decision criteria (see Figures \ref{fig:Card_Estimand_bounds} and \ref{fig:CardTableOpt}).

\subsubsection{Encoding of variables}
\label{app: NLSYM}
We bin $L$'s continuous covariates as follows. The parents' last education level is separated into 3 categories: "High-School", "Undergraduate" and "Graduate"; IQ scores are separated into 5 categories "50-69", "70-89", "90-109", "110-129" and "130-149"; Age is separated into 2 categories: "14-19" and "19-24". Missing variables are dealt with using mean imputation following \citet{Card1993} and \citet{Wang2018}.
\subsubsection{Models}
\label{app : NLSYM models}
To estimate the $p_{ya.z}(l)$, we fitted multinomial regressions of $Y \times A$ on $L$ conditional on $Z = 0$ and $Z = 1$. For the $(l,a')$-CATE and $(l,a')$-value functions, we fitted models for $\mathbb{E}(Y | L)$, $\mathbb{E}(Y | L, A = 0)$ and $\mathbb{E}(Y | L, A = 1)$ using three logistic regressions of $Y$ on $L$ using the full data, conditioning on $A = 0$, and conditioning on $A = 1$. We also use a logistic regression of $A$ on $L$ to estimate the propensity score $P(A = 1 | L = l)$.

\subsubsection{$(L,A,Z)$-optimal regimes}
\label{app: NLSYM analysis conditional on Z}
Recall the $(L,A,Z)$-regimes defined in Definition \ref{def: (L,A,Z)-optimal regime}. Under Assumptions \ref{As:Unconf} and \ref{As:Excl} of the main text, $Y^a \independent Z | L \Rightarrow \mathbb{E}(Y^a | L, Z) = \mathbb{E}(Y^a | L)$ \citep{StensrudSarvet2022}. Then, we have an analogous statement to Lemma \ref{lemmma1} in the main text conditional on $Z$.
\begin{proposition}
Under Assumptions \ref{ass: cconsistency}, \ref{As:Unconf} and \ref{As:Excl} of the main text, and assuming $P(A = a' | L, Z) > 0$ with probability 1,  
\begin{align}
 \mathbb{E}(Y^{a} \mid A = a', L=l, Z = z)
 & =  \begin{cases}
                    \mathbb{E}(Y \mid A = a', L=l, Z = z),   & \text{if } a = a' , \\
                    \frac{\mathbb{E}(Y^{a} \mid L=l) - \mathbb{E}(Y \mid A = a, L=l, Z = z)P(A =a \mid L = l, Z = z)   }{P(A =a' \mid L = l, Z = z)},  & \text{if }  a \neq a'. 
                \end{cases} \label{eq: lemma}
\end{align}
\end{proposition}
Then, analogously to Proposition \ref{Equiv_PartId}, we can sharply bound $\mathbb{E}(Y^a \mid A = a', L=l, Z = z)$ using the sharp \cite{BalkePearl1997} bounds.
\begin{proposition}
    Under Assumptions \ref{As:Unconf} and \ref{As:Excl} of the main text, the $(L,Z)$-optimal regime is equal to the $L$-optimal regime.
    \label{cor: (L,Z) = (L)}
\end{proposition}

Recall the covariates $l \in \mathcal{L}$ of Section \ref{sec:NLSYM data analysis}. We do a similar analysis, but conditional on the instrument $Z$, omitting the $(L,Z)$-optimal regime as it is equal to the $L$-optimal regime.
\begin{figure}
    \centering
    \includegraphics[scale = 0.4]{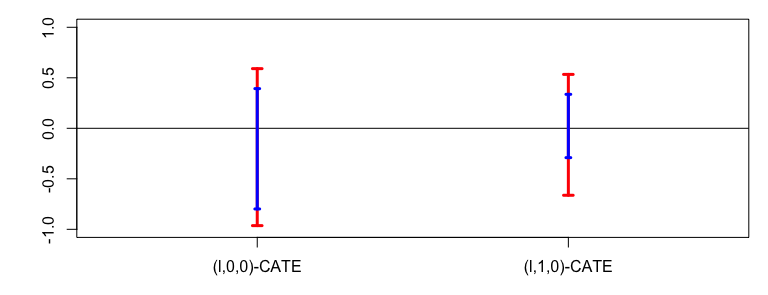}
    \caption{Bounds on the $(L,A,0)$-CATEs in the NLSYM analysis.}
    \label{fig:Card_CombinedCATEZ0}
\end{figure}

\begin{figure}
    \centering
    \includegraphics[scale = 0.4]{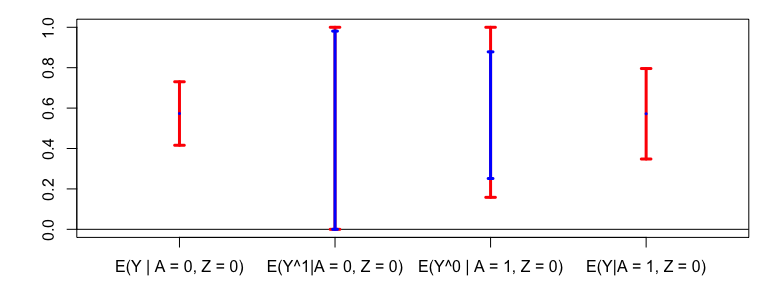}
    \caption{Estimated bounds in the NLSYM analysis $(Z = 0)$.}
    \label{fig:Card_Estimand_boundsZ0}
\end{figure}

The bounds on the $(L,A,0)$-CATEs cover zero (see Figure \ref{fig:Card_CombinedCATEZ0}), but a decision maker could justify decisions based on explicit criteria \citep{Cui2021Individualized} using the bounds for the $\mathbb{E}(Y^a | L = l, Z = 0)$ (see Figures \ref {fig:Card_Estimand_boundsZ0} and \ref{fig:CardTableOptZ0}). The decision criteria do not change when we condition on $Z = 0$ (see Figure \ref{fig:CardTableOptZ0}).
\begin{figure}
    \centering
    \begin{tabular}{ |p{4cm}||p{3cm}|p{3cm}|p{3cm}|p{2cm}|p{2cm}| }
 \hline
 Decision criteria  & $A = 0$ & $A = 1$\\
 \hline
 Maximax(Optimist)    & $a = 1$ & $a = 0$\\
 Maximin(Pessimist)  & $a = 0$ & $a = 1$\\
 Minimax(Opportunist)  & $a = 0$ & $a = 0$\\
 Healthcare decision making($a = 0$ baseline) & $a = 0$ & $a = 0$\\
 \hline
\end{tabular}
    \caption{Outcome of several decision criteria in the NLSYM analysis, $Z = 0$ case.}
    \label{fig:CardTableOptZ0}
\end{figure}

\begin{figure}
    \centering
    \includegraphics[scale = 0.4]{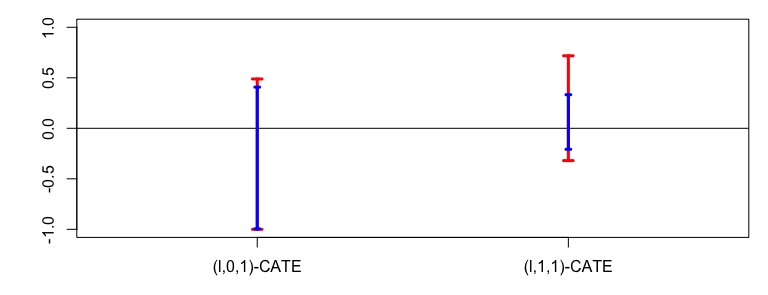}
    \caption{Bounds on the $(L,A,1)$-CATEs in the NLSYM analysis ($Z = 1$).}
    \label{fig:Card_CombinedCATEZ1}
\end{figure}
\begin{figure}
    \centering
    \includegraphics[scale = 0.4]{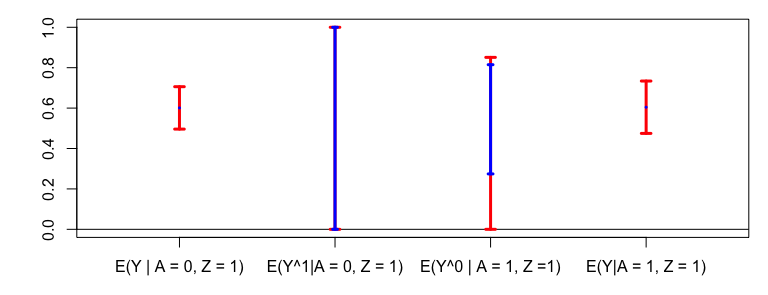}
    \caption{Bounds in the NLSYM analysis ($Z = 1$).}
    \label{fig:Card_Estimand_boundsZ1}
\end{figure}

The bounds of the $(L,A,1)$-CATEs also cover zero (see Figure \ref{fig:Card_CombinedCATEZ1}), but the $(l,1,1)$-optimal opportunist makes a different decision than the $(l,1)$ and $(l,1,0)$-optimal opportunist using bounds for $\mathbb{E}(Y^a | L = l, Z = 1)$ (see Figures \ref {fig:Card_Estimand_boundsZ1} and \ref{fig:CardTableOptZ1}).
\begin{figure}
    \centering
    \begin{tabular}{ |p{4cm}||p{3cm}|p{3cm}|p{3cm}|p{2cm}|p{2cm}| }
 \hline
 Decision criteria & $A = 0$ & $A = 1$\\
 \hline
 Maximax(Optimist)   & $a = 1$ & $a = 0$\\
 Maximin(Pessimist) & $a = 0$ & $a = 1$\\
 Minimax(Opportunist)  & $a = 0$ & $\color{red} a = 1$\\
 Healthcare decision making($a = 0$ baseline)  & $a = 0$ & $a = 0$\\
 \hline
\end{tabular}
    \caption{Outcome of several decision criteria in the NLSYM analysis ($Z = 1$). The differences between the $(L,A,0)$ and $(L,A)$-optimal decision criteria are highlighted in red.}
    \label{fig:CardTableOptZ1}
\end{figure}

\subsection{Effect of influenza vaccination on hospitalizations} \label{sec: Influenza regime bounds example}
We studied the effect of a seasonal influenza vaccine ($A$) on an indicator for absence of flu-related hospital visits ($1-Y$); that is, we study outcomes under regimes where the seasonal influenza vaccine is optimally assigned based on observed covariates $L$, which include the individual's age, gender, and indicators for chronic obstructive pulmonary disease, diabetes mellitus, heart disease, severe renal failure, and chronic liver failure. The $(L,A)$-optimal regimes also use the doctor's intended treatment recommendation $A$. Our data were derived from a trial that randomly assigned physicians to receive reminder messages to encourage inoculation of patients \citep{McDonald1992}. We let the randomly assigned messages be instruments ($Z = 1$ if the patients' physician received the reminder and $Z = 0$ if they did not) and, like others \citep{Hirano2000,Imbens2014,levis2023covariateassisted}, we ignore clustering by physician since we do not know which patients were seen by which physicians. 

In vaccine and infectious disease settings, it is possible that outcomes of some patients may depend on other vaccinated patients in the population, violating the Stable Unit Treatment Value Assumption (SUTVA) of \citet{Rubin1980}. We maintain the SUTVA assumption here analogously to previous data analyzes \citep{Hirano2000,Imbens2014,levis2023covariateassisted}, but the results should be interpreted in light of this potential violation.

\begin{figure}
    \centering
    \includegraphics[scale = 0.55]{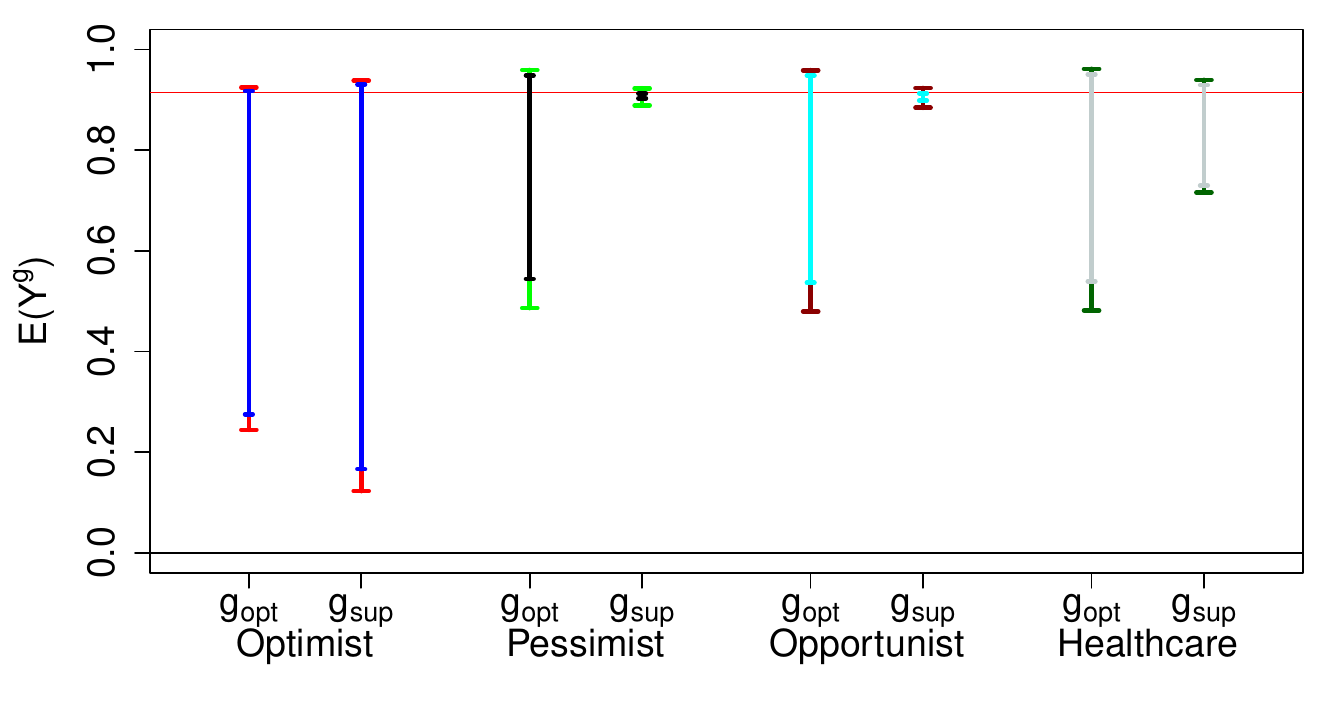}
    \caption{Bounds of value functions in the influenza example.}
    \label{fig:Influenza_regime_bounds}
\end{figure}

We computed Balke-Pearl bounds and the confidence intervals, using the same type of estimators as in Section \ref{sec:NLSYM data analysis}. The differences between  $\hat{g}_{\textbf{sup}}$ and $\hat{g}_{\textbf{opt}}$ are more prominent than in the NLSYM data analysis in Figure \ref{fig:Card_regime_bounds} (see Figure \ref{fig:Influenza_regime_bounds}).

The estimated $(L,A)$-optimal regimes have substantially narrower bounds compared to the estimated $L$-optimal regimes, without imposing additional assumptions (Figure \ref{fig:Influenza_regime_bounds}). This observation is not surprising: the estimated $(L,A)$-optimal regime might be identical to the implicit regime that was implemented in the observed data for several $l \in \mathcal{L}$. When  the estimated $(L,A)$-optimal regime and the implicit regime are identical, their conditional value functions are point identified as $\mathbb{E} ( Y^{g_{\textbf{sup}}} | A = a) = \mathbb{E}(Y^a | A = a) = \mathbb{E}(Y | A = a)$ by Assumption \ref{ass: cconsistency}. 

The observed regime has an estimated value function of $0.915$ (CI: $(0.904, 0.925)$). Our analysis suggests that only minor improvements can (possibly) be achieved by replacing the observed regime with an algorithmic decision regime that only uses the covariates $L$ considered in this analysis, and the doctor's natural treatment value; in particular, the bounds on the estimated $(L,A)$-optimal regimes suggest that the algorithm will follow the doctor's natural treatment value for many values of $A$ and $L$, and that the doctors used information not encoded in $L$.

\subsection{Non-binary $A$}
\label{app: Non-binary A}
When $A$ has more than two categories, identification of the $(L,A)$-optimal regimes from observed data requires more care. Let $A \in \mathcal{A}$, where $\mathcal{A}$ is a countable set with more than two elements. The law of total expectation does not in general permit us to identify all of the $\mathbb{E}(Y^a | A = a', L = l) $, for $a \neq a'$, from $\mathbb{E}(Y^a | L = l)$ as in Lemma \ref{lemmma1}. Indeed, we have one equation with  $|\mathcal{A}| - 1 > 1$ unknowns. However, we can still use the law of total expectation to bound $\mathbb{E}(Y^a | A = a', L = l)$ using $\mathbb{E}(Y^a | L = l)$ as follows.
\begin{proposition}
    Let $A \in \mathcal{A}$ and $a,a' \in \mathcal{A}$ where $a \neq a'$. Then under Assumptions \ref{ass: cconsistency} and \ref{ass: positivity}, 
    \begin{align*}
        &\frac{\mathbb{E}(Y^a | L = l) - \mathbb{E}(Y | A = a, L = l)P(A = a| L = l) + P(A = a | L = l) + P(A = a' | L = l) - 1}{P(A = a' | L = l)}\\
        &\leq \mathbb{E}(Y^a | A = a', L = l)\\
        &\leq \frac{\mathbb{E}(Y^a| L = l) - \mathbb{E}(Y | A = a, L = l)P(A = a| L = l)}{P(A = a' | L = l)}.
    \end{align*}
    \label{Prop: Discrete A}
\end{proposition}
\begin{proof}
    Using the law of total expectation,
    \begin{align*}
        \mathbb{E}(Y^a |  L = l) = \sum_{a''} \mathbb{E}(Y^a | A = a'', L = l)P(A = a'' | L = l)
    \end{align*}
    Then solving for $\mathbb{E}(Y^a | A = a', L = l)$, using Assumptions \ref{ass: cconsistency} and \ref{ass: positivity}, the bounds $0 \leq \mathbb{E}(Y^a | A = a', L = l) \leq 1$, and $1 - P(A = a | L = l) - P(A = a' | L = l) = \sum_{a'' \neq a,a'} P(A = a'' | L = l)$, we obtain the bounds in the statement of the proposition.
\end{proof}
\begin{corollary}
    For bounds $\mathbf{L}(l) \leq \mathbb{E}(Y^a | L = l) \leq \mathbf{U}(l)$ of width $\rho(l) = \mathbf{U}(l) - \mathbf{L}(l)$, we can construct bounds on $\mathbb{E}(Y^a | A = a', L = l)$ of width $\min(\frac{\rho(l) + 1 - P(A = a | L = l) - P(A = a' | L = l)}{P(A = a', L = l)},1)$
\end{corollary}
\begin{proof}
    Using Proposition \ref{Prop: Discrete A}, 
    \begin{align*}
        &\max(0,\frac{\mathbf{L}(l) - \mathbb{E}(Y | A = a, L = l)P(A = a| L = l) + P(A = a | L = l) + P(A = a' | L = l) - 1}{P(A = a' | L = l)})\\
        &\leq \mathbb{E}(Y^a | A = a', L = l)\\
        &\leq \min(1,\frac{\mathbf{U}(l) - \mathbb{E}(Y | A = a, L = l)P(A = a| L = l)}{P(A = a' | L = l)})
    \end{align*}
    are valid bounds of $\mathbb{E}(Y^a | A = a', L = l)$. Subtracting the upper bound from the lower bound gives us the formula for the width given in the statement of the Corollary.
\end{proof}
We cannot reproduce the sharpness result of Proposition \ref{Equiv_PartId} as we bound $\mathbb{E}(Y^a \mid A = a', L = l)$ using Proposition \ref{Prop: Discrete A} and using the bounds on $\mathbb{E}(Y^a |  L = l)$. 

\subsection{Bound widths}
\label{app: Bound widths}
\subsubsection{Bound widths and observed performance}
We are interested in covariates $l$ for which the optimal and superoptimal regimes match the observed treatment, which implies the value function is point identified and hence tightens the bounds of the marginal value function of the regime.
\begin{proposition}
    Let $w^a(l)$ be the width of the bounds of $\mathbb{E}[Y^a | L = l]$ and $\rho^a(a',l) = I[a \neq a']w^a(l)/P(A = a'|L = l)$ be the width of the induced bounds of $\mathbb{E}[Y^a | A = a', L = l]$. The induced bounds of $\mathbb{E}[Y^{g^{\textbf{sup}}}]$ will be tighter than those of $\mathbb{E}[Y^{g^{\textbf{opt}}}]$ if and only if
    \begin{align*}
        \sum_{a',l} P(L = l) \bigg((I[g_{\textbf{sup}}(a',l) \neq a'] - I[g_{\textbf{opt}}(l) \neq a']) w^{1-a'}(l) - I[g_{\textbf{opt}}(l) = a']w^{a'}(l) \bigg) < 0.
    \end{align*}
    \label{Prop: Bound widths and expert performance}
\end{proposition}

\begin{proof}

    The induced bounds of $\mathbb{E}[Y^{g^{\textbf{sup}}}]$ have width
    \begin{align*}
        w^{g_{\textbf{sup}}} &:= \sum_{a',l} P(A = a', L= l)\rho^{g_{\textbf{sup}}(a',l)}(a',l)\\
        &= \sum_{a',l} P(L = l) I[g_{\textbf{sup}}(a',l) \neq a']w^{1-a'}(l).
    \end{align*}
    However, the induced bounds of $\mathbb{E}[Y^{g^{\textbf{opt}}}]$ have width
    \begin{align*}
        w^{g_{\textbf{opt}}}&:= \sum_l P(L = l) w^{g_{\textbf{opt}}}(l)\\
        &= \sum_l P(L = l) \sum_{a'} I[g_{\textbf{opt}}(l) = a']w^{a'}(l) + I[g_{\textbf{opt}}(l) \neq a'] w^{1-a'}(l)\\
        &= \sum_{a',l} P(L = l) (I[g_{\textbf{opt}}(l) = a']w^{a'}(l) + I[g_{\textbf{opt}}(l) \neq a'] w^{1-a'}(l)).
    \end{align*}
    Hence, 
    \begin{align*}
        w^{g_{\textbf{sup}}} - w^{g_{\textbf{opt}}} &= \sum_{a',l} P(L = l)\bigg((I[g_{\textbf{sup}}(a',l) \neq a'] - I[g_{\textbf{opt}}(l) \neq a']) w^{1-a'}(l)\\
        &- I[g_{\textbf{opt}}(l) = a']w^{a'}(l) \bigg).
    \end{align*}
\end{proof}
\begin{corollary}
    The width of the induced bounds of $\mathbb{E}[Y^{g_{\textbf{sup}}}]$ decreases as the set $\{(a',l): g_{\textbf{sup}}(a',l) \neq a'\}$ gets smaller, and the difference $w^{g_{\textbf{sup}}} - w^{g_{\textbf{opt}}}$ decreases as $\{(a',l): g_{\textbf{sup}}(a',l) = a' \neq g_{\textbf{opt}}(l) \}$ gets larger; the superoptimal regime's bounds will be smaller when the observed regime outperforms the optimal regime. However, the difference increases as $\{(a',l): g_{\textbf{sup}}(a',l) \neq a'\}$ becomes larger; the superoptimal regime's bounds will become larger if the observed regime underperforms.
\end{corollary}
\subsubsection{Decision criteria and bound widths}
To fix ideas about the width of the bounds, consider first the setting where a decision criterion coincides with the observed regime.  The bounds on the value function of this regime will have width zero, i.e., it is point identified.  The superoptimal conventionality criterion, as proposed in Section \ref{sec: Decision-making criteria}, never deviated from the observed regime in Figure \ref{fig:ICU_regime_bounds}. More broadly, the width of the bounds under the superoptimal decision criteria depends on how frequently the criterion coincides with the observed regime, whose value function conditional on $A$ and $L$, $\mathbb{E}[Y | A, L]$, is point identified. When the bounds of $\mathbb{E}[Y^a | L = l]$ are wide, the bounds of the $(l,1-a)$-CATE will often cover zero, that is, $\mathbf{L}(l,1-a) < 0 < \mathbf{U}(l,1-a)$. In this setting, the superoptimal opportunist chooses the observed regime $1-a$ if $|\mathbf{U}(l,1-a)| > |\mathbf{L}(l,1-a)|$ for $a = 0$ and $|\mathbf{U}(l,1-a)| < |\mathbf{L}(l,1-a)|$ for $a = 1$; the healthcare decision maker chooses their baseline regimes, $a = 0$; the superoptimal pessimist will choose to implement the observed treatment $1-a$; and the superoptimal optimist will choose to implement $a$, although $\mathbb{E}[Y^a | A = 1-a, L = l]$ is not point identified. If bounds of $\mathbb{E}[Y^a | L = l]$ are wide, we expect the value function under a superoptimal pessimist regime to have narrow bounds. Similarly, we expect the superoptimal optimist to have wide bounds, and the superoptimal opportunist to have narrower bounds than the optimist, but larger width than the pessimist, because when $\mathbf{L}(l,1-a) < 0 < \mathbf{U}(l,1-a)$ the superoptimal opportunist can choose $1-a$ or $a$. Such trends are seen in Figure \ref{fig:ICU_regime_bounds}.

\subsection{Proofs}
\label{app: proofs}
\subsubsection{Proof of Proposition \ref{Equiv_PartId}}
\label{app: Proof of Equiv_PartId}
Consider $a \neq a^*$. Using Assumption \ref{ass: cconsistency}, Lemma \ref{lemmma1} and the law of total expectation,
\begin{align}
    \mathbb{E}(Y^{a} - Y^{a^*} \vert A = a, L = l) & = \mathbb{E}(Y^a \vert A = a, L = l) - \mathbb{E}(Y^{a^*} \vert A = a, L = l) \nonumber\\
    & = \mathbb{E}(Y \vert A = a, L = l) \nonumber\\
    & - \frac{\mathbb{E}(Y^{a^*}\vert L = l) - \mathbb{E}(Y \vert A = a^*, L = l)P(A = a^*\vert L = l)}{P(A = a \vert L = l)} \nonumber\\
    & = \frac{\mathbb{E}(Y \vert L = l) - \mathbb{E}(Y^{a^*}\vert L = l)}{P(A = a\vert L = l)},
    \label{eq: CATE id}
\end{align}
which shows that we can derive bounds of $\mathbb{E}(Y^{a} - Y^{a^*} \vert A = a, L = l)$ from bounds on $\mathbb{E}(Y^{a^*}\vert L = l)$ . The expression for the width of the bounds also follows immediately from Equation \eqref{eq: CATE id}.

To show that the induced bounds are sharp, we argue by contradiction. Consider sharp bounds on $\mathbb{E}(Y^a | L = l)$ of width $\phi(l)$. Denote by $\rho(a^*,l) = \rho(a^*,l) = \phi(l)/P(A = a^* | L = l)$ the width of the bounds on $h_{\mathbf{sup}}(a^*,l)$ constructed from the sharp bounds as in Proposition \ref{Sign_prop}. Suppose we have also have sharper bounds of width $\varrho(a^*,l) < \rho(a^*,l)$ on $h_{\mathbf{sup}}(a^*,l)$. Then by the previous statements in the proposition, the bounds on $h_{\mathbf{opt}}(a^*,l)$ of width $\varrho(a^*,l)$ induce bounds on $\mathbb{E}(Y^a | L = l)$ that are of width $P(A = a | L = l) \cdot \varrho(a^*,l) < P(A = a | L = l) \cdot \rho(a^*,l) = \phi(l)$. This contradicts the sharpness of the bounds on $h_{\mathbf{sup}}(a^*,l)$ and proves our result.
\subsubsection{Proof of Proposition \ref{Sign_prop}}
The proof follows from the law of total expectation,
\begin{align*}
    h_{\mathbf{opt}}(l) = h_{\mathbf{sup}}(a',l)P(A = a' | L = l) + h_{\mathbf{sup}}(1-a',l) P(A = 1-a' | L = l).
\end{align*}
By positivity,
\begin{align*}
    h_{\mathbf{sup}}(1-a',l) = \frac{h_{\mathbf{opt}}(l) - h_{\mathbf{sup}}(a',l)P(A = a' | L = l)}{P(A = 1 - a' | L = l)}.
\end{align*}
All the results in the proposition then follow, using an appropriate choice of bounds.

\subsubsection{Proof of Proposition \ref{width_prop}}
By the definitions of $\mathbf{U}(l)$ and $\mathbf{L}(l)$, $\mathbf{U}(l) - \mathbf{L}(l) = \mathbf{U}_1(l) - \mathbf{L}_1(l) + \mathbf{U}_0(l) - \mathbf{L}_0(l)$, and
\begin{align*}
    \mathbf{L}(l,1) & = \mathbb{E}(Y | A = 1, L = l) - \frac{\mathbf{U}_0(l) - \mathbb{E}(Y | A = 0, L = l) P(A = 0| L = l)}{P(A = 1 | L = l)}\\
    & = \frac{\mathbb{E}(Y | L = l) - \mathbf{U}_0(l)}{P(A = 1 | L = l)}.
\end{align*}
Similarly, 
\begin{align*}
    \mathbf{U}(l,1) = \frac{\mathbb{E}(Y | L = l) - \mathbf{L}_0(l)}{P(A = 1 | L = l)}.
\end{align*}
Hence,
\begin{align*}
    \phi(1,l) = \mathbf{U}(l,1) - \mathbf{L}(l,1) = \frac{\mathbb{E}(Y | L = l) - \mathbf{L}_0(l)}{P(A = 1 | L = l)} - \frac{\mathbb{E}(Y | L = l) - \mathbf{U}_0(l)}{P(A = 1 | L = l)} = \frac{\mathbf{U}_0(l) - \mathbf{L}_0(l)}{P(a = 1 | L = l)}.
\end{align*}
Again, we can see
\begin{align*}
    \mathbf{U}(l,0) & = \frac{\mathbf{U}_1(l) - \mathbb{E}(Y | L = l)}{P(A = 0|L = l)}, \text{ and}\\
    \mathbf{L}(l,0) & = \frac{\mathbf{L}_1(l) - \mathbb{E}(Y | L = l)}{P(A = 0| L = l)},
\end{align*}
which implies
\begin{align*}
    \phi(0,l) = \mathbf{U}(l,0) - \mathbf{L}(l,0) = \frac{\mathbf{U}_1(l) - \mathbb{E}(Y | L = l)}{P(A = 0| L = l)} - \frac{\mathbf{L}_1(l) - \mathbb{E}(Y | L = l)}{P(A = 0| L = l )} = \frac{\mathbf{U}_1(l) - \mathbf{L}_1(l)}{P(A = 0 | L = l)}.
\end{align*}
And then,
\begin{align*}
    \omega(l) = P(A = 1 | L = l) \phi(1,l) + P(A = 0 | L = l) \phi(0,l).
\end{align*}
A similar argument gives an analogous convex relationship of the lower and upper bounds of the CATEs, from which we can deduce the results of the corollaries of this proposition. In particular,
\begin{align*}
    \mathbf{L}(l) &= P(A = 0| L = l) \mathbf{L}(l,0) + P(A = 1| L = l) \mathbf{L}(l,1),\\
    \mathbf{U}(l) &= P(A = 0| L = l) \mathbf{U}(l,0) + P(A = 1| L = l) \mathbf{U}(l,1).
\end{align*}

\subsubsection{Proof of Corollary \ref{cor: sign id}}
Corollary \ref{cor: sign id} follows immediately from Proposition \ref{thm:proof eff infl} in the main text. The influence function of $\mathbb{E}(Y)$ is
\begin{align*}
    Y - \mathbb{E}(Y).
\end{align*}
Using differentiation rules, we find that the influence function of $\mathbb{E}(Y ) - \mathbb{E}(Y^a)$ is
\begin{align*}
    Y - \mathbb{E}(Y) - \psi_1^{\text{eff}},
\end{align*}
which is the formula given in the statement of the corollary.

\subsubsection{Proof of Proposition \ref{prop: superopt healthcare with obs base}}
The second inequality follows directly from the definition of $g_{A-\text{opt}}$.

If $ \mathbb{E}[Y | L = l] < \mathbf{L}_a(l)$,
\begin{align*}
     \mathbb{E}[Y | L = l] &<P(A = a | L = l)\mathbb{E}[Y | A = a, L = l]\\
    &+ P(A = 1 - a| L = l) \frac{\mathbf{L}_a(l) - P(A = a | L = l)\mathbb{E}[Y | A = a, L = l]}{P(A = 1 - a  | L = l)}.
\end{align*}
Hence,
\begin{align}
     \mathbb{E}[Y | A = 1 - a, L = l] &< \frac{\mathbf{L}_a(l) - P(A = a | L = l)\mathbb{E}[Y | A = a, L = l]}{P(A = 1 | L = l)}, \text{ which implies}  \label{eq: lower bound of super conv better}\\
      0 &< \mathbf{L}(l, 1-a). \nonumber
\end{align}
Hence, whenever the optimal conventionality criterion improves on the observed regime, the superoptimal conventionality criterion does the same.  This proves the statement of Proposition \ref{prop: superopt healthcare with obs base}.

The right hand side of Equation \eqref{eq: lower bound of super conv better} is the induced lower bound of $\mathbb{E}[Y^a | A = 1-a, L = l]$, see online Appendix \ref{app: Sharpness}. Hence, the lower bound of the superoptimal conventionality criterion will be greater or equal to it's optimal counterpart.

\subsubsection{Proof of Proposition \ref{Criteria_Equiv}}
We first consider $A = 1$. Recall that,
\begin{align*}
    \mathbf{L}(l,1) & = \frac{\mathbb{E}(Y | L = l) - \mathbf{U}_0(l)}{P(A = 1 | L = l)}
\end{align*}
and so $\mathbf{L}(l,1) > 0$ if and only if $\mathbf{U}_0(l) < \mathbb{E}(Y | L = l)$.\\
Then, the optimistic decision criterion is given by
\begin{align*}
I(\mathbb{E}(Y | A = 1, L = l) )> \frac{\mathbf{U}_0(l) - \mathbb{E}(Y | A = 0, L = l)P(A = 0 | L = l)}{P(A = 1 | L = l)} \})
    = I(\mathbb{E}(Y | L = l) )> \mathbf{U}_0(l)\}),
\end{align*}
which is equivalent to the healthcare decision criterion in the statement of Proposition \ref{Criteria_Equiv}.

Now consider $A = 0$,
\begin{align*}
    \mathbf{L}(l,0) & = \frac{\mathbf{L}_1(l) - \mathbb{E}(Y | L = l)}{P(A = 0| L = l)},
\end{align*}
 hence $\mathbf{L}(l,0) > 0$ if and only if $\mathbf{L}_1(l) > \mathbb{E}(Y | L = l)$. However in this case, the pessimistic decision criterion is given by
\begin{align*}
&I(\{\mathbb{E}(Y | A = 0, L = l) < \frac{\mathbf{L}_1(l) - \mathbb{E}(Y | A = 1, L = l)P(A = 1 | L = l)}{P(A = 0 | L = l)} \})\\
    &= I(\mathbb{E}(Y | L = l) < \mathbf{L}_1(l)),
\end{align*}
which is then equivalent to the healthcare decision criterion. 
\subsubsection{Proof of Proposition \ref{Mixed_and_Opp}}
Our proof follows the proof of Theorem 5.1 in \cite{Cui2021Individualized}. Let $p$ be the probability of taking $A = 1$ when $L = l$ and $A = a'$. The optimal randomized strategy is then given by the following optimization problem,
\begin{align*}
    \min_{p} \max([1 - p]\max\{\mathbf{U}(l, a'),0\}, p \max\{-\mathbf{L}(l, a'),0\}),
\end{align*}
which is minimized when
\begin{align*}
    p^* & = \begin{cases} 1 & \mathbf{L}(l, a') > 0\\
    0 & \mathbf{U}(l, a') < 0\\
    \frac{\mathbf{U}(l, a')}{\mathbf{U}(l, a') - \mathbf{L}(l, a')} & \mathbf{L}(l, a') < 0 < \mathbf{U}(l, a')
    \end{cases}.
\end{align*}
Then, the worst-case regret is bounded by
\begin{align*}
    \begin{cases}
    0 & \mathbf{U}(l, a') < 0 \text{ or } \mathbf{L}(l, a') > 0\\
    -\frac{\mathbf{L}(l, a') \mathbf{U}(l, a')}{\mathbf{U}(l, a') - \mathbf{L}(l, a')} & \mathbf{L}(l, a') < 0 < \mathbf{U}(l, a')
    \end{cases}.
\end{align*}
If we define $g_{\textbf{stoch}}$ to be the strategy that outputs $1$ with probability $p^*$, the regret is bounded by
\begin{align*}
    &\mathbb{E}(Y^{g_{\textbf{sup}}} | A = a', L = l) - \mathbb{E}(Y^{g_{\textbf{stoch}}} | A = a', L = l)\\
    &\leq \mathbb{E}\left(  -\frac{\mathbf{L}(l, a') \mathbf{U}(l, a')}{\mathbf{U}(l, a') - \mathbf{L}(l, a')} I\{\mathbf{L}(l, a') < 0 < \mathbf{U}(l, a') \}\right).
\end{align*}
If we use the classical opportunistic decision criterion, we get a worst-case regret (when $L = l$) of
\begin{align*}
    \min(\max\{\mathbf{U}(l, a'),0\}, \max\{-\mathbf{L}(l, a'),0\}).
\end{align*}
However,
\begin{align*}
    -\frac{\mathbf{L}(l, a') \mathbf{U}(l, a')}{\mathbf{U}(l, a') - \mathbf{L}(l, a')} < \min \{ -\mathbf{L}(l, a'), \mathbf{U}(l, a') \}, \text{ if $\mathbf{L}(l, a') < 0 < \mathbf{U}(l, a')$}.
\end{align*}
In this sense, the mixed strategy is better than the opportunistic strategy in expectation.

\subsubsection{Proof of Proposition \ref{prop: One-step convergence}}
We follow a similar argument to the proof of Theorem \ref{thm: Levis} from \citet{levis2023covariateassisted}.

We first prove the following lemma.
\begin{lemma}
    $\mathbb{E}_P[\Psi^{\text{eff}}_{d^{\Psi}}] = 0$.
\end{lemma}
\begin{proof}
    \begin{align*}
        \mathbb{E}_P[\Psi^{\text{eff}}_{d^{\Psi}}] = \mathbb{E}_P[\sum_{a'\in \{0,1\}}\mathbb{E}_P[\Psi^{\text{eff}}_{d^{\Psi}} | A = a',L]P(A = a'| L = l)].
    \end{align*}
    Let $p(a|l) = P(A = a | L = l)$. Then, by Proposition \ref{thm:proof eff infl},
    \begin{align*}
        &\mathbb{E}_P[\Psi^{\text{eff}}_{d^{\Psi}} | A = a',L] \\
        &= \mathbb{E}_P\bigg[\frac{1}{p(1-a'|L)^2} \Bigg(\underbrace{\mathbb{E}_P[\psi_1^{\text{eff}}(1-a',L)]}_{= 0}p(a'|L)\\
        &- \psi_1(1-a'|L) \underbrace{(p[a'|L] - p(a'|L)}_{ = 0}\\
        &- \bigg(\frac{1}{p(1-a'|L)}\underbrace{(\mathbb{E}_P[YI(A = 1-a') | L] - p(1-a'|L)\mathbb{E}[Y  \mid  A = 1-a', L])}_{= 0} p(1-a'|L)\\
        & + \mathbb{E}[Y  \mid  A = 1-a',L]\underbrace{(p(1-a'|L)- p(1-a'|L))}_{=0}\bigg) \cdot p(a'|L) \\
        &+ \mathbb{E}[Y  \mid  A = 1-a', L] p(1-a'|L) \underbrace{(p(a'|L) - p(a'|L))}_{=0} \Bigg)\bigg]\\
        &= 0.
    \end{align*}
\end{proof}
Hence,
\begin{align*}
    \mathbb{P}_n (\hat{\Psi}_{\hat{d}^{\Psi}} + \hat{\Psi}_{\hat{d}^{\Psi}}^{\text{eff}}) - \Psi &= (\mathbb{P}_n - P)(\hat{\Psi}_{\hat{d}^{\Psi}} + \hat{\Psi}_{\hat{d}^{\Psi}}^{\text{eff}} - \Psi_{d^{\Psi}} + \Psi_{d^{\Psi}}^{\text{eff}}) + P(\hat{\Psi}_{\hat{d}^{\Psi}} + \hat{\Psi}_{\hat{d}^{\Psi}}^{\text{eff}}  - \Psi_{d^{\Psi}} - \Psi^{\text{eff}}_{d^{\Psi}})\\
    &+ (\mathbb{P}_n - P)\Psi.
\end{align*}
The following simple algebra result will be useful for the derivation of the following results.
\begin{lemma}
For $u,v,\hat{u}, \hat{v} \in \mathbb{R}$, $\hat{u}v - u\hat{v} = (\hat{u} - u)v + u(v - \hat{v})$.
\label{lemma: error difference expr}
\end{lemma}
We now prove the following intermediate results.
\begin{lemma}
Under the conditions of Proposition \ref{prop: One-step convergence},
\begin{align*}
    \sqrt{n}(\mathbb{P}_n - P)\Psi \to \mathcal{N}(0,\sigma^2),
\end{align*}
for some $0< \sigma^2 < \infty$.
\end{lemma}
\begin{proof}
This is a direct application of the central limit theorem.
\end{proof}

\begin{lemma}
    Under the conditions of Proposition \ref{prop: One-step convergence},
    \begin{align*}
        P(\hat{\Psi}_{\hat{d}^{\Psi}} + \hat{\Psi}_{\hat{d}^{\Psi}}^{\text{eff}} - \Psi_{d^{\Psi}} - \Psi_{d^{\Psi}}^{\text{eff}}) = o_P(n^{-1/2}).
    \end{align*}
\end{lemma}
\begin{proof}
    \begin{align*}
        &\hat{\Psi}(a,l)  - \Psi(a,l)\\
        &= I(a = g(a,l))\bigg(\hat{\mathbb{E}}[Y | A = a, L = l] - \mathbb{E}[Y | A = a, L = l] \bigg)\\
        &+ I(a \neq g(a,l))\bigg(\frac{\hat{\psi}_1(a,l) - \hat{\mathbb{E}}[Y | A = 1-a, L = l]\hat{P}(A = 1-a| L = l)}{\hat{P}(A = a | L = l)}\\
        &- \frac{\psi_1(a,l) - \mathbb{E}[Y | A = 1-a, L = l]P(A = 1-a| L = l)}{P(A = a | L = l)} \bigg)\\
        &= I(a = g(a,l))\bigg(\hat{\mathbb{E}}[Y | A = a, L = l] - \mathbb{E}[Y | A = a, L = l] \bigg)\\
        &+ I(a \neq g(a,l))\bigg((\hat{\psi}_1(a,l) - \hat{\mathbb{E}}[Y | A = 1-a, L = l]\hat{P}(A = 1-a| L = l)) P(A = a | L = l)\\
        &- (\psi_1(a,l) - \mathbb{E}[Y | A = 1-a, L = l]P(A = 1-a| L = l)) \hat{P}(A = a | L = l)\bigg)P(A = a | L = l)^{-1}\\
        &= I(a = g(a,l))\bigg(\hat{\mathbb{E}}[Y | A = a, L = l] - \mathbb{E}[Y | A = a, L = l] \bigg)\\
        &+ I(a \neq g(a,l))\bigg((\hat{\psi}_1(a,l) - \psi_1(a,l))P(A = a | L = l)\\
        &+ \psi_1(a,l)(P(A = a | L = l) - \hat{P}(A = a | L = l))\\
        &+ (\mathbb{E}[Y | A = 1-a, L = l]P(A = 1-a| L = l)\\
        &- \hat{\mathbb{E}}[Y | A = 1-a, L = l]\hat{P}(A = 1-a| L = l))P(A = a | L = L)\\
        & + \mathbb{E}[Y | A = 1-a, L = l]P(A = 1-a| L = l)(\hat{P}(A = a | L = l) - P(A = a | L = l))\bigg)\\
        &= I(a = g(a,l))\bigg(\hat{\mathbb{E}}[Y | A = a, L = l] - \mathbb{E}[Y | A = a, L = l] \bigg)\\
        &+ I(a \neq g(a,l))\bigg((\hat{\psi}_1(a,l) - \psi_1(a,l))P(A = a | L = l)\\
        &+ \psi_1(a,l)(P(A = a | L = l) - \hat{P}(A = a | L = l))\\
        &+ \bigg((\mathbb{E}[Y | A = 1-a, L = l] - \hat{\mathbb{E}}[Y | A = 1-a, L = l])P(A = 1-a| L = l)\\
        &+\hat{\mathbb{E}}[Y | A = 1-a, L = l](P(A = 1-a| L = l) - \hat{P}(A = a | L = l))\bigg)P(A = a | L = L)\\
        & + \mathbb{E}[Y | A = 1-a, L = l]P(A = 1-a| L = l)(\hat{P}(A = a | L = l) - P(A = a | L = l))\bigg).\\
    \end{align*}
    Furthermore, 
    \begin{align*}
        &\hat{\Psi}^{\text{eff}}(a,l) - \Psi^{\text{eff}}_{d^{\Psi}}(a,l)\\
        &= \frac{1}{\hat{p}(1-g(a,l)|l)^2 p(1-g(a,l)|l)^2}\\
        &\cdot\bigg[\hat{\psi}_1^{\text{eff}}(a,l)\hat{p}(1-g(a,l)|l)p(1-g(a,l)|l)^2 - \psi_1^{\text{eff}}(a,l)p(1-g(a,l))\hat{p}(1-g(a,l)|l)^2\\
        & -\hat{\psi}_1(a,l)(I(A = 1- g(a,l)) - \hat{p}(1-g(a,l)|l))p(a = 1-g(a,l)|l)^2\\
        &+ \psi_1(a,l)(I(A = 1- g(a,l)) - p(1-g(a,l)|l))\hat{p}(a = 1-g(a,l)|l)^2\\
        &- \bigg(I[a = g(a,l)](y - \hat{\mathbb{E}}[Y  \mid  A = g(a,l) , L = l])\\
    & + \hat{\mathbb{E}}[Y  \mid  A = g(a,l), L = l](I[ a= g(a,l)] - \hat{p}(g(a,l)|l))\bigg) \cdot \hat{p}(1-g(a,l)|l)p(1-g(a,l)|l)^2 \\
    &- \bigg(I[a = g(a,l)](y - \mathbb{E}[Y  \mid  A = g(a,l), L = l])\\
    & + \mathbb{E}[Y  \mid  A = g(a,l), L = l](I[ a= g(a,l)] - p(g(a,l)|l))\bigg) \cdot p(1-g(a,l)|l)\hat{p}(1-g(a,l)|l)^2 \\
    &+ \hat{\mathbb{E}}[Y  \mid  A = g(a,l), L = l] \hat{p}(g(a,l) |l)p(a = 1-g(a,l)|l)^2\\
    &\cdot (I[A = 1-g(A,L)] - \hat{P}(A = 1-g(A,L)| L = l))\\
    & - \mathbb{E}[Y  \mid  A = g(A,L), L = l ] P( A = g(A,L) | L = l ) \\
    &\cdot(I[A = 1-g(A,L)] - P(A = 1-g(A,L) | L = l))\cdot\hat{p}(1-g(a,l)|l)^2\bigg].
    \end{align*}
    Using Lemma \ref{lemma: error difference expr}, and the boundedness conditions of Proposition \ref{prop: One-step convergence}, we can express all of the above as sums of differences between estimators and true values multiplied by bounded terms. Hence, by the convergence conditions of Proposition \ref{prop: One-step convergence}, we obtain the desired result.
\end{proof}

\begin{lemma}
    Under the conditions of Proposition \ref{prop: One-step convergence},
    \begin{align*}
        (\mathbb{P}_n - P)(\hat{\Psi} - \Psi) = o_P(n^{-1/2}).
    \end{align*}
\end{lemma}
\begin{proof}
    The proof follows by computing the differences as in the proofs above, except it requires that the estimators of $\mathbb{E}[Y | A,L]$ and $p(a|l)$ are $P$-Donsker as in the statement of the conditions of Proposition \ref{prop: One-step convergence}, not just $\sqrt{n}$-convergent, as we consider the operator $(\mathbb{P}_n - P)$ instead of $P$.
\end{proof}

\begin{remark}
    We can relax the Donsker conditions of Proposition \ref{prop: One-step convergence} if we do sample splitting for the estimators and the empirical average, and that the estimators of $\mathbb{E}[Y | A,L]$ and $p(a|l)$ converge at $\sqrt{n}$-rate, see for example \citet{Chernozhukov2018}.
\end{remark}

\end{document}